\documentclass[3p,twocolumn,authoryear,11pt]{elsarticle}

\usepackage{graphicx}

\usepackage{amssymb}

\usepackage{color}
\usepackage{natbib}
\usepackage{latexsym}
\biboptions{compress,comma}
\journal{Planetary Space Science, accepted for publication}

\begin{document}

\begin{frontmatter}


\title{Radio astronomy with the Lunar Lander: opening up the last unexplored frequency regime.}

\author[run,st]{Marc Klein Wolt\corref{cor1}}
\ead{M.KleinWolt@astro.ru.nl}
\author[run]{Amin Aminaei}
\author[lesia]{Philippe Zarka}
\author[sron]{Jan-Rutger Schrader}
\author[astron]{Albert-Jan Boonstra}
\author[run]{Heino Falcke}

\cortext[cor1]{Corresponding author}

\address[run]{Astronomical Institute, Radboud University Nijmegen, Heijendaalseweg 135, 6525 AJ Nijmegen, The Netherlands}
\address[st]{Science \& Technology, Olof Palmestraat 14, 2616 LR Delft, The Netherlands}
\address[lesia]{Laboratoire d'\'{e}tudes Spatiales et d'Instrumentation en Astrophysique (LESIA) Observatoire de Paris, Section de Meudon 5, Place Jules Janssen, 92195 Meudon Cedex France}
\address[sron]{Netherlands Institute for Space Research (SRON), Sorbonnelaan 2, 3584 CA, Utrecht, The Netherlands}
\address[astron]{Netherlands Institute for Radio Astronomy (Astron), Oude Hoogeveensedijk 4, 7991 PD, Dwingeloo, The Netherlands}

\begin{abstract}
The moon is a unique location in our solar system and provides important information regarding the exposure to free space that is essential for future human space exploration to mars and beyond. The active broadband (1 kHz$-$100 MHz) tripole antenna now envisaged to be placed on the European Lunar Lander located at the Lunar South Pole allows for sensitive measurements of the exosphere and ionosphere, and their interaction with the Earths magnetosphere, solar particles, wind and CMEs and studies of radio communication on the moon, that are essential for future lunar human and science exploration. In addition, the lunar South pole provides an excellent opportunity for radio astronomy. Placing a single radio antenna in an eternally dark crater or behind a mountain at the south (or north) pole would potentially provide perfect shielding from man-made radio interference (RFI), absence of ionospheric distortions, and high temperature and antenna gain stability that allows detection of the 21 cm wave emission from pristine hydrogen formed after the big bang and into the period where the first stars formed. A detection of the 21 cm line from the moon at these frequencies would allow for the first time a clue on the distribution and evolution on mass in the early universe between the Epoch of Recombination and Epoch of Reionization (EoR). Next to providing a cosmological breakthrough, a single lunar radio antenna would allow for studies of the effect of solar flares and Coronal Mass Ejections (CMEs) on the solar wind at distances close to earth (space weather) and would open up the study of low frequency radio events (flares and pulses) from planets such as Jupiter and Saturn, which are known to emit bright (kJy$-$MJy) radio emission below 30 MHz \citep{JF2009}. Finally, a single radio antenna on the lunar lander would pave the way for a future large lunar radio interferometer; not only will it demonstrate the possibilities for lunar radio science and open up the last unexplored radio regime, but it will also allow a determination of the limitations of lunar radio science by measuring the local radio background noise.
\end{abstract}


\end{frontmatter}


\section{Introduction}
\label{intro}

The European Lunar Lander ELL mission is an exploratory ESA mission that aims at investigating the conditions on the Lunar surface for future human missions and science projects. The primary objective of the Lunar Lander is to prove ESA's capabilities of performing a soft precision landing in the moon, and once on the moon the Lunar Lander will perform a number of scientific studies. The L-DEPP (Lunar Dust Environment and Plasma Package) package is one of the proposed instrument packages to be placed on the Lunar Lander and it consists of three instruments: a radio antenna (Lunar Radio eXperiment - LRX), dust camera and Langmuir probe to measure the lunar electromagnetic fields and waves, lunar dust and plasma, respectively. The radio antenna was also a consequence of our early proposal for ESA's Exploration architecture studies to build a lunar low-frequency radio array (Lunar LOFAR) where it was intended as a path-finder instrument proving the capabilities of low-frequency astronomy (below $\sim$ 30~MHz).

The ELL will land on one of the predefined landing sites on the Lunar South Pole. The South Pole was chosen based on arguments related to power (sunlight) and communication (Earth visibility) but also as it has not been explored before and offers unique opportunities for future human exploration \citep[e.g. resource mining, see][]{gardini2011}. For the work presented here we will assume the South Pole location as a fixed parameter. The LRX offers unique scientific opportunities but it should be noted that these science cases are not driving the ELL mission.

Here we would like to present a concept design for, and review the relevance of a single tripole radio antenna on board the Lunar Lander on the a Lunar South Pole location, not only from the lunar exploration perspective but also highlighting the unique opportunities such an radio antenna offer for low frequency radio astronomy and cosmology. For a more detailed description of the science cases that support a space-based or lunar low frequency radio antenna we refer to \citep{JF2009}.

\section{Opportunities provided by the Lunar Lander}
\label{lunarlander}

The moon is a unique and still relatively little studied place in our solar system. From ESA's exploratory perspective the moon is the next step after the ISS for pursuing exploration of the universe, and provides the ideal location to test human habitation and scientific exploration in preparation for a future mars mission. However, the extreme conditions on the moon have to be taken into account for any mission to the surface. The moon is known for having a significant dust layer and, in combination with the absence of a significant atmosphere, provides a unique interface between a dusty surface and free space. In addition, the moon is subjected to large temperature variations (caused by night and day variations), is exposed to comic ray and solar wind plasma and interacts with the Earth magnetotail. All this has a large impact on the lunar ionosphere and plasma and affects conditions below and above the lunar surface. For example, it is obvious that levitating dust plays a crucial and potentially hazardous role for surface operations. However, how does the dust get there and where does it go? In fact, the dusty surface plasma will not end just a few meters above the surface. Since a few decades we know, for example, that the moon has an exosphere. It is generally believed that the ions, which make up this exosphere, are generated at the Moon's surface by interaction with solar UV photons, plasma in the Earth's magnetosphere, or micrometeorites \citetext{see \citealt{RB72}, \citealt{BFHea75} and references therein}. Apart from direct in-situ measurements this exosphere has been seen with radio observations, revealing a variable plasma frequency in the $\>100$~kHz regime during daytime.

One of the main scientific objectives is to investigate the interplay between the dust, the electromagnetic fields and lunar plasma on the one hand and the effects of the earth's magnetic fields, solar light and plasma, and cosmic ray and (micro-) meteorites. The synergy between the three instruments proposed to comprise L-DEPP provides a unique opportunity to study these effects. As explained above, the solar UV and X ray photons are predicted to create a positively charged surface and a dusty plasma sheet on the day side, through the combined effects of photo electron emissions and low conductivity of the lunar regolith. Pick-up ions and solar wind electrons should create a lunar ionosphere that is more extended on the night side; a night side that has a negatively charged surface as an interface to the ionosphere that gradually becomes a near perfect vacuum inside the lunar wake (see also Fig.~\ref{fig:wake} and \ref{fig:daynight}). The generated electric fields should often dominate the Lorentz force, breaking MHD plasma equilibrium and enabling levitation of dust and transport of dust and plasma across the terminator. It is far from clear, though, how all these processes work in detail. Clearly, placing the three L-DEPP instruments on the South Pole of the Moon would allow for detailed in-situ measurements and remote sensing of the effects, which should provide significantly more insight into the underlying processes

The radio antenna as part of the L-DEPP package will study the variations in the variations in the lunar exosphere, the lunar radio background, and the effect of high-energy cosmic rays that interact with the Lunar surface, predominantly in combination with measurements from the dust camera and Langmuir probe. In that respect the radio antenna offers the unique capability to perform remote sensing measurements, that will complement the in-situ capabilities of the L-DEPP package. In addition, the LRX will investigate the use of radio communication on the moon. The low frequency radiation is able to penetrate through the lunar regolith or crater rims; it can be reflected from the lunar surface or even the lunar exosphere. And finally, using the radio antenna as passive ground-penetrating radar, the reflected radiation provides information about the lunar surface and sub-surfaces. 

The LRX radio antenna plays a crucial role in investigating the lunar conditions for human exploration, but furthermore offers unique opportunities for low frequency radio astronomy: a single antenna placed on the moon would be a pathfinder mission for radio astronomy and address pending cosmological issues. Placing a broad-band and low frequency ($\sim$kHz$-100$MHz) radio antenna on the moon would be the demonstration and site-testing mission required to make large low-frequency radio antenna systems in space possible. However, LRX would also detect the Askaryan effect from cosmic rays and make a serious attempt at detecting the global dark ages and EOR signal. The main limitation for the latter is the unknown interference situation at a polar landing site, where the Earth would set only occasionally during the mission. A future large radio synthesis antenna with many more individual nodes and baselines of a few kilometers would be required to perform detailed cosmological studies such as the EoR tomography and the determining the power spectrum of the 21-cm line at redshifts of $30-50$, as well as for instance the study of high redshift galaxies and quasars and fossil radio galaxies.

\subsection{South Pole Location}
\label{sp}

\begin{figure}
\centering
\includegraphics[width=\columnwidth]{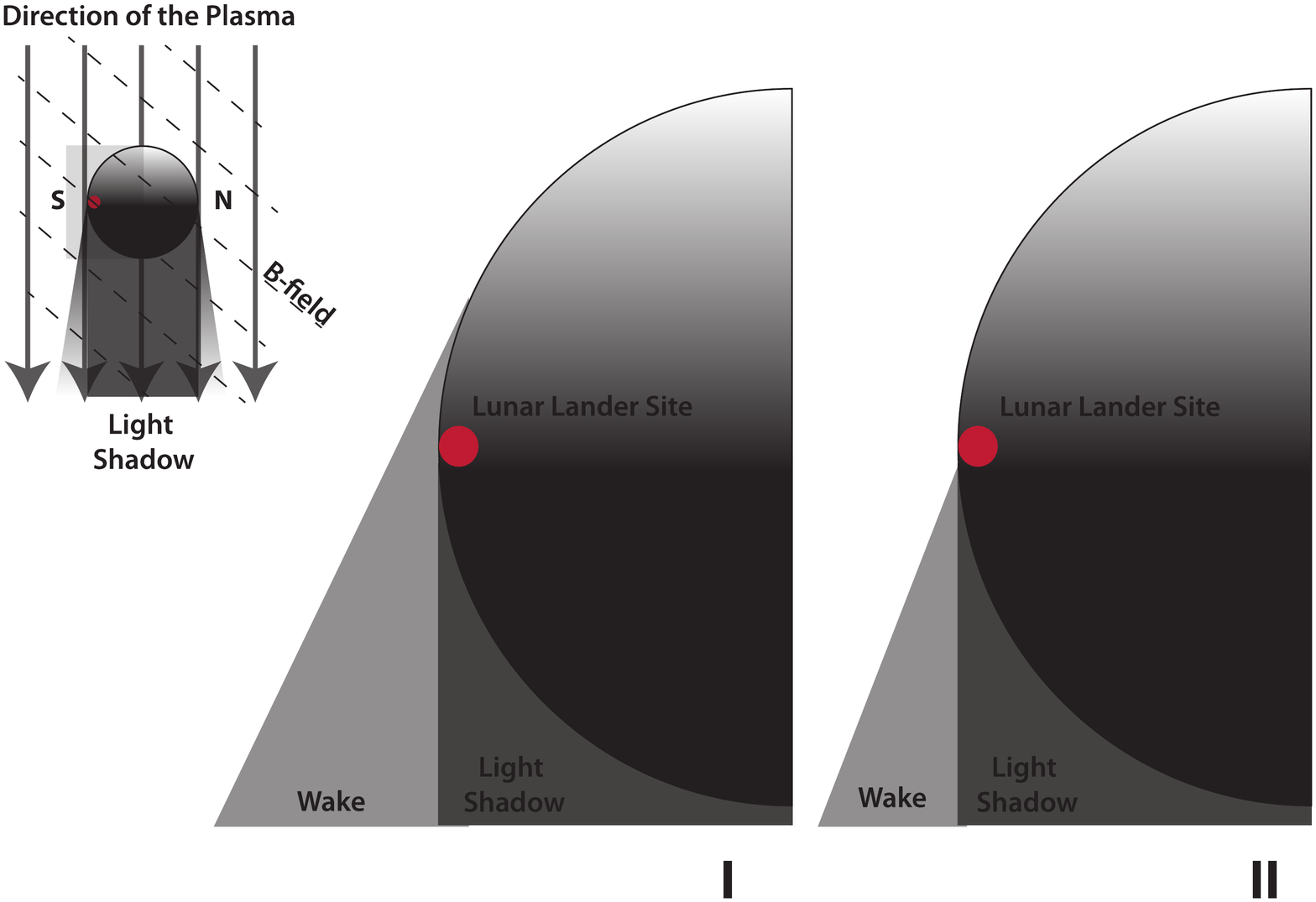}
\caption{The changing lunar wake boundary at the location of the Lunar Lander site at the South Pole of the Moon. The red dot corresponds to the location of the Lunar Lander. The image on the top-left shows the overall situation of the Moon embedded in the interplanetary magnetic field (IMF) and solar wind plasma, the light shadow and the wake. The images on the right-hand side correspond to the light-grey rectangular shown in the top-left figure. The changes in the plasma of the sun and the rotation of the Moon change the location of the boundary wake and place the Lunar Lander inside and outside of the wake.\label{fig:wake}}
\end{figure}

A strong argument in favor of measurements on the lunar South Pole is the possibility to study the effects of the solar wind and the Earth's magnetosphere on the lunar wake. As explained above, the Moon is embedded in the solar wind plasma which creates a wake on the lunar night side. The wake is basically a bubble in the supersonic solar wind plasma and is one of the best vacuums in our local space environment: measurements by the WIND spacecraft showed that the electron density in the wake is less than 0.01cm$^{3}$ \citep[][and see Fig.~\ref{fig:wake}]{op96}. The LRX can observe the Lunar wake through the plasma turbulence that is created at the boundary wake and result in emission at both low and high frequencies.  

Figure~\ref{fig:wake} shows the shape of the lunar wake and the location of the boundary on the lunar surface in the South Pole area. The shape and location can change under the influence of changes in the solar wind plasma and the Earth's magnetic field, to the extreme case when the wake completely disappears; that is, when the Moon is completely embedded in the magnetosphere. This change in shape and location is shown as I and II in the figure: a change in the wake will cause the Lunar Lander to move in and out of the wake. Hence, the South Pole location would provide an excellent location for studying the impact of the solar wind plasma and the Earth's magnetosphere on the local Lunar environment (on the South Pole): the changes in the wake have a direct impact on the density of the local plasma and its plasma frequency and even a fixed location would still allow for plasma and ionospheric measurements in and outside the wake.

\begin{figure}
\centering
\includegraphics[width=\columnwidth]{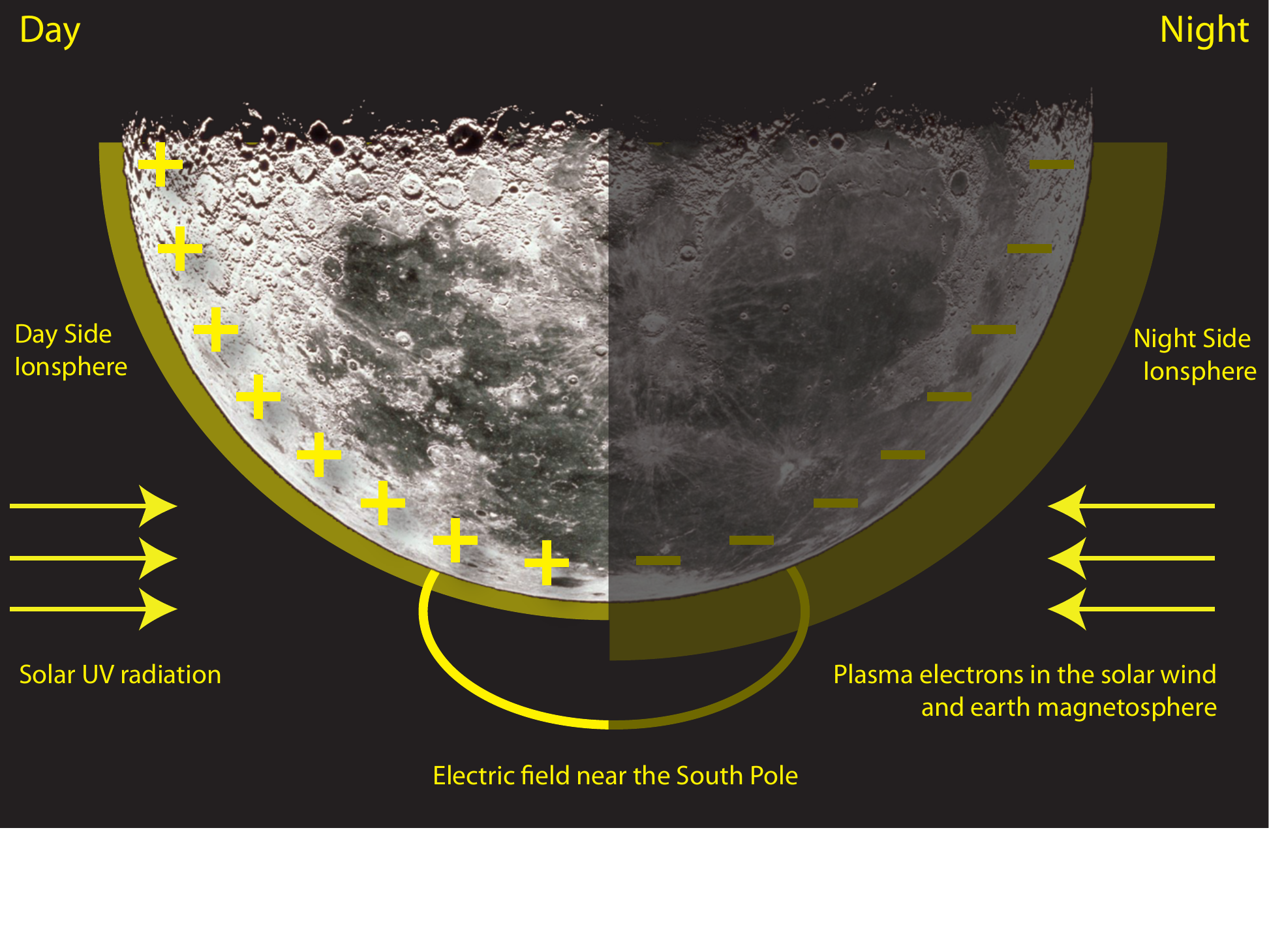}
\caption{Day and night difference on the lunar South Polar location: solar UV radiation causes the lunar surface to be 
positively charged on the day-side (a few Volts, extending up to $\sim1$meter in height), on the night-side 
the interaction with plasma electrons (from the solar wind and the earth's magnetosphere) causes the 
surface to become negative ($\sim100$V, extending up to $\sim1$ km). This causes a strong electric field on the 
south and north pole. In addition the moon is constantly exposed to micrometeorites and cosmic rays.\label{fig:daynight}}
\end{figure}

As presented in Figure~\ref{fig:daynight}, the lunar South Pole is an ideal location as it provides a continuous view of the lunar ionosphere above the dark and sunlit side of the Moon. This is the main advantage over a day-side or night-side location were only one instance of the ionosphere can be observed. Understanding the effects of both the solar UV photons as well as the interaction with plasma electrons are crucial for the formation of the ionosphere, the electric and magnetic characteristics of the Moon and the formation of the lunar dusty plasma sheet. 

It is not within the scope of this work to argue for or against specific landing sites, e.g. North Pole or equator, instead we provide an overview of the scientific possibilities that the LRX provide at the Lunar South Pole. What makes the ELL unique and the South Pole and ideal location, compared to for instance space-based alternatives, is the possibility to perform in-situ measurements on the Lunar environment and the possibility of performing low-frequency radio observations, in particular the 21-cm cosmology. We will present a more quantitative analysis on the landing site requirements including an analysis of the expected RFI levels, in a forthcoming paper. 

\section{Lunar Exploration with the LRX}
\label{exploration}

From the Lunar Exploration point of view there are a number of interesting science issues on which the LRX can provide an essential contribution. In this section we present these in some more detail.

\subsection{Lunar Ionosphere}

The issue of a possible lunar ionosphere has arisen from the intriguing dual frequency radar measurements from the Luna 22 spacecraft. They indicated that a rather dense plasma surrounds the Moon up to about 50 km from the surface \citep{Vyshlov1976}. Electron number densities as large as 2000 cm$^{-3}$ were inferred from the measured plasma column contents, which suggested that the Moon indeed has a dense ionosphere that could interact readily with the surrounding solar wind or geo-tail plasma. The mass loading of the geo-tail by such an ionosphere, if true, could be enough to cause large disturbances and even cause auroral storms closer to Earth. However, a reinterpretation of the Luna 22 data by Bauer \citep{Bauer1996}, in light of the previously neglected very effective ion pick-up transport process expected to occur near the Moon by the solar wind, suggests that no ionospheric layer of any significance is to be expected around the Moon. Instead, the Luna 22 measurements are interpreted as a result of a photo electron layer near the surface, in support of the ALSEP observations \citep{RO72}. This controversy is confirmed by more recent radio occultation measurements using SMART-1, Cassini, Venus Express \citep{pluch2008} and the SELENE radio experiment \citep{ima2008}. \cite{pluch2008} find electron density numbers in agreement with in-situ measurements by the US Apollo missions, while \cite{ima2008} find electron densities in the order of 1000 cm$^{-3}$ on the day side, much higher then the predicted 1 cm$^{-3}$. More recently, \cite{goto11} suggests a new method for examining the lunar ionosphere by using the propagation of AKR radio emission. Their first results using this new method shows no evidence for a high electron density layer near the lunar surface. Clearly, the present proposal to make careful \emph{in-situ} measurements of the lunar ionosphere using LRX in combination with the other instruments in the L-DEPP package, will provide better insights and confirm or deny the conclusions arrived at earlier.

Little is yet known on the lunar ionosphere but that is likely to change rather soon due to the presence of the two ARTEMIS spacecraft in lunar orbits and the launch of the Lunar Atmosphere and Dust Environment Explorer (LADEE) in 2013. The ARTEMIS spacecraft that were sent to the Moon stem from them magnetospheric THEMIS mission, which consisted of four satellites. Although not originally intended as a Moon mission, the two spacecraft are fully equipped with modern plasma instruments and plan to go as low as 15 km above the surface during several orbits, and, as its name implies, the main objective of the LADEE mission is to characterize the lunar atmosphere and dust environment. However, although data from the ARTEMIS and LADEE missions will be of extremely high value for the lunar science community and for future L-DEPP science planning, both missions employ equatorial orbits and cannot perform any measurements above the poles, neither are they stationary: only a polar lander mission can accomplish the science goals set up by L-DEPP, which makes the European Lunar Lander unique.

It is currently believed that the lunar ionosphere is composed of ions, released from the lunar surface through interactions with the solar UV radiation in a process known as ion pick-up, and electrons in the solar wind and the Earth's magnetosphere \citep[see for instance][]{well1962}. Its heavier constituents are likely due to impacts of high-energy cosmic rays and micrometeorites. The density profile and physical extent of the lunar ionosphere are still unknown. In addition, large variations in the electron density are known to occur. As shown in Fig.~\ref{fig:daynight} the physical extent of the ionosphere is different on the day side and night side, mainly caused by the strong effect of the solar UV radiation (ion pick-up) on the day-side and the interaction with plasma electrons from the solar wind and earth magnetic field on the night-side. The lunar ionosphere and its variations caused by the presence or absence of the Earth's magnetosphere, gradual changes in the solar wind speed and interplanetary shocks due to solar flares and Coronal Mass Ejections (CME), and impacts of cosmic rays and micrometeorites have never been studied in-situ. However, on the lunar day-side a variable plasma frequency greater than 100 kHz has been revealed by radio observations. Note that the presence of (charged) dust moving through the ionosphere of the moon could effect the local radio background noise as their discharges are detected as they interfere with the ELL and the LRX, see for instance \cite{stubbs2007}. Determining the size, speed and concentration of the dust in the lunar ionosphere is being addressed by the Dust Camera which is part of the L-DEPP package.

To obtain high sensitivity, a radio telescope requires long term stability. In order to assess radio astronomy from the Moon, it is therefore important that the LRX antenna can sense narrowband interferences \citep[narrow-band features are expected to be present in the lunar plasma, in the same way as they are predicted to occur in the Earth's atmosphere, see][]{allan92} as well as any fast temporal changes. These would be flagged and later removed from the astronomical data. However, one man's noise is another man's signal. Narrowband emissions as well as transients are important ingredients in the lunar radio background. Although the plasma frequency in the solar wind is only about 10 Hz, which is difficult to measure with a short antenna, it could potentially be as high as 3 MHz on the surface, if the plasma density measurements made by Apollo ALSEP were correct. The plasma frequency would show up as locally generated, rather narrow band, Langmuir wave emission, or as an abrupt cut-off in the in the remote radio spectrum. It is also possible that nonlinear plasma waves on the harmonics of the plasma frequency are emitted. Furthermore, Auroral kilometric radiation (AKR) from Earth is rather broadband and can go up to 1 MHz. The AKR can be very strong and should be detectable even with a relatively short antenna. There are also the strong coherent broadband radio bursts from the Sun and from other radio planets than Earth, the Jovian decametric radiation in particular, which can last up to one hour. While not desirable for radio astronomy, the non-thermal radio emission from Earth and Jupiter are most useful since they would appear as mostly circularly polarized point sources. This makes it possible to perform direction finding from a single location using three mutually orthogonal dipole or monopole antennas \citep[][Bergman et al, Swedish Patent No. 512 219 (2000), US Patent No. 6,407,702 (2002)]{CKB00}, which in turn makes it possible to estimate the total electron content in the direction of the source. Ionospheric tomography could be carried out as the Sun or the radio planets travel across the lunar sky. Note that, given that objects like planets and the Sun will only be visible a few degrees above the local horizon (depending on the exact landing site) due to the South Pole location,the Ionospheric variations may possibly interfere with the actual scientific observations of these objects.

\subsection{Lunar impacts}

Monitoring the temporal radio background could potentially be used to directly measure impacts of high-speed dust particles and micro-meteorites. During the Cassini mission ring crossings, for instance, dust-particles hitting the spacecraft were detected as spikes by the radio antennas. The lunar surface provides an even larger collecting area. However, how this plays out on the Moon is uncertain. It would be surprising, however, if a sudden impact, creating a very hot plasma cloud, would not also lead to an electromagnetic pulse that could be detected at low frequencies. One can ask a similar but more speculative question about possible discharges. If indeed, the lunar surface is strongly charged and the charging is related to solar radiation, strong potential differences should exist between dark and bright areas. Note that the LRX would be able to determine the Direction of Arrival (DOA) with a few degrees accuracy (see Section\ref{planets}). Again, the provision of charged particles (e.g. through an impact or spontaneously) could lead to static discharges that would be detectable as a radio hiss anywhere in the $1-100$ MHz regime. This effect has never been studied in any detail and is a bit venturing into unknown territory. After all, there has never been a modern broad-band receiver on the Moon, which could have detected and localized such events. Also, one needs a large surface without atmosphere to see this in the first place and hence other planets are of no use either. If charging of the surface plays a role, such effects should be strongest near the terminator and near the poles and would have been much less pronounced during Apollo sortie missions.

If a micro-meteoroid hits the lander or the regolith nearby, a plasma cloud is formed instantaneously, producing a strong EMP with a broad frequency spectrum that can be measured by LPX and LRX. The electrical effects of meteoroids on spacecraft systems are not well known and this provides a valuable in situ measurement. The pulse width and amplitude of the radio emission contains information about the plasma formation and expansion caused by the impact, which can be used to infer the mass and velocity of the meteoroid. More specifically it can address the a) charge production b) length scale c) plasma temperature d) ionization efficiency and e) plasma expansion speed \citep{close2010}. It is also estimated that radio emissions have a peak at low to medium frequencies so the LPX probes and LRX antennas would use the LF and MF bands of LRX to sense and attempt to localize the events.

\subsection{Lunar radio communication}

Low frequency radio transmission could also be considered for over-the horizon emergency communication for future surface sorties, i.e. if a manned or unmanned rover accidentally or purposely comes in a situation where there is no line of sight communication to a base station/lander and direct to Earth or satellite communication are impossible or have failed. In this case a low-frequency beacon may be desirable. After all, low-frequency radiation can penetrate deep through the regolith or crater rims, may refract over the surface, or may even be reflected by a lunar exosphere. On the other hand, shielding of radio emission by mountains or crates may actually be required by future radio telescopes. 

These effects have been modeled to some degree but at low frequencies they are not well understood, since these models suffer from the fact that the wavelengths become of the scale of tens of meters to actually kilometers and hence are much larger than any sample ever brought back to Earth \citep[see for instance][]{taka03}. Moreover, the dusty plasma sheet close to the surface will likely also play a significant role for the propagation of radio waves parallel to the surface. Radio propagation effects on the lunar surface at low frequencies cannot be probed with Earth-based low-frequency radars. Even if they could penetrate the ionosphere, which is sometimes possible even below 10 MHz, it would be impossible to tell if any detectable radio wave propagation effects were due to the lunar environment or caused by Earth's ionosphere, the latter being most likely. Hence, studies of radio wave propagation on the lunar surface have to be carried out on site. One method to measure the radio wave propagation effects on site is to observe the setting of a strong radio source, such as Jupiter or the Earth below the horizon. Such observations will be carried out routinely to perform ionospheric tomography but as the radio source sets, propagation effects due to the lunar geography should start to play a dominating role.

\section{Low Frequency Radio Astronomy with the LRX}
\label{lfastro}

\begin{figure}
\centering
\includegraphics[width=\columnwidth]{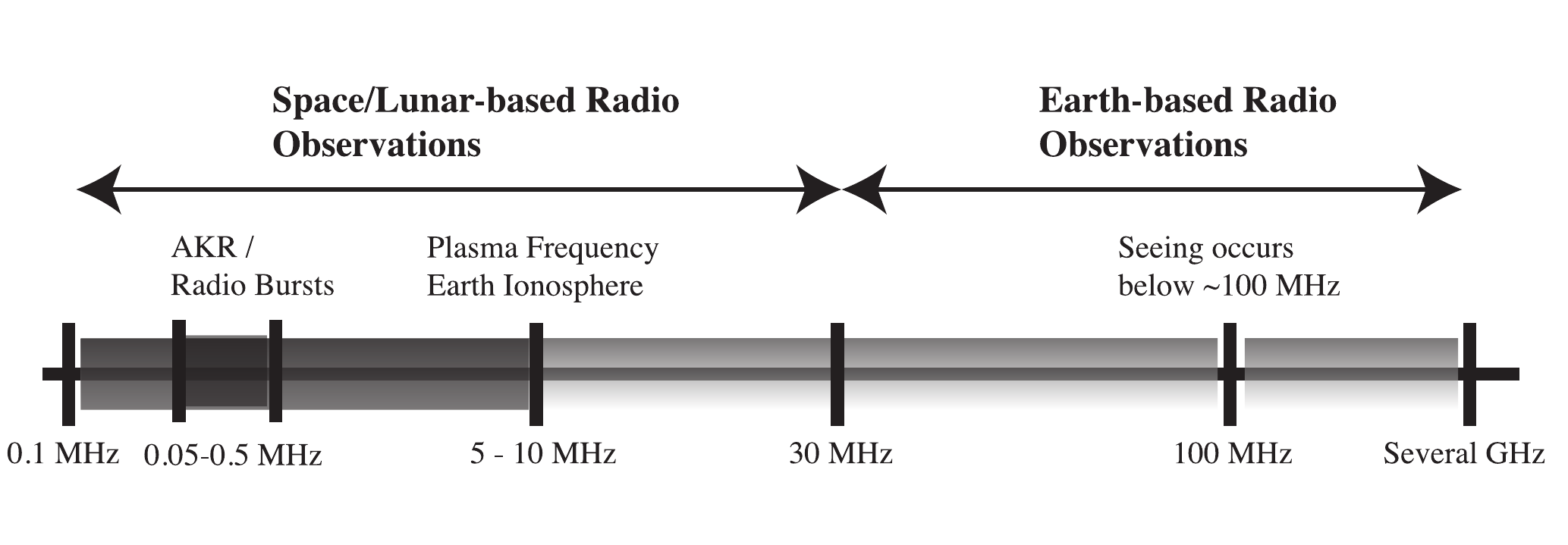}
\caption[]{Typical space/lunar- and earth-based frequency regimes.  \label{fig:freqregime}}
\end{figure}

The currently planned ground-based telescopes will provide a serious advance in radio astronomy and extend the accessible frequencies to the widest range possible from the ground. As illustrated by Fig.~\ref{fig:freqregime}, Earth's ionosphere gives rise to "radio seeing'', making ground-based radio imaging of the sky difficult at frequencies below $\sim100$ MHz. At even longer wavelengths, in the $10-30$ MHz range, one often encounters severe ionospheric turbulence and Faraday rotation, and below $\sim10$ MHz the ionospheric cut-off blocks incoming radio waves completely, prohibiting any radio astronomical observations whatsoever. Observing just above the cut-off, i.e., between $\sim10-30$ MHz is sometimes possible but to obtain any decent images requires especially favorable geomagnetic and ionospheric conditions, which usually does not last for very long time periods and therefore limits the sensitivity of the observations. The frequency range below the cut-off can only be observed from space. Hence, the dominant low-frequency (long-wavelength) regime for which ground-based telescopes are being designed is at frequencies above 30 MHz (wavelengths below 10 m). Even above this frequency, there are windows with strong radio frequency interferences (RFI), e.g. FM radio broadcasting stations in the $80-120$ MHz band, which cannot be used for radio astronomy.

The best resolution achieved so far in the range below $\sim30$ MHz is on the scale of around 5 degrees, but more typically of order tens of degrees. This compares rather unfavorably to the milli-arcsecond resolution that can be routinely obtained in very long baseline interferometry (VLBI) at higher radio frequencies. Hence, the low-frequency Universe is the worst charted part of the radio spectrum, and perhaps even of the entire electromagnetic spectrum.  By today, only two kinds of maps of the sky have been made at frequencies below 30 MHz. The first are maps of a part of the southern sky near the Galactic center such as those obtained by \citet{CaneWhitham77,EllisMendillo87,CaneErickson01} from Tasmania. These have angular resolutions ranging from 5 to 30 degrees. The second kind are maps obtained by the RAE-2 satellite \citep{NB78} with angular resolution of 30 degrees or worse. None of these maps show individual sources other than diffusive synchrotron emission of the Galaxy, nor do they cover the entire sky. Even though the new LOFAR telescope will attempt to improve on this at frequencies above 10 MHz, lower frequencies remain off limit. 

To improve this situation and to overcome these limitations, space-based low-frequency telescopes are required for all observations below the ionospheric cutoff \citep{Weiler87,WJSea88,KassimWeiler90}. This is also true for a significant part of the seeing-affected frequency range above the cutoff frequency where high-resolution and high-dynamic range observations are required, such as imaging of 21-cm emission of neutral hydrogen in the very early Universe \citep{CHL07}.

\begin{figure}
\centering
\includegraphics[width=\columnwidth]{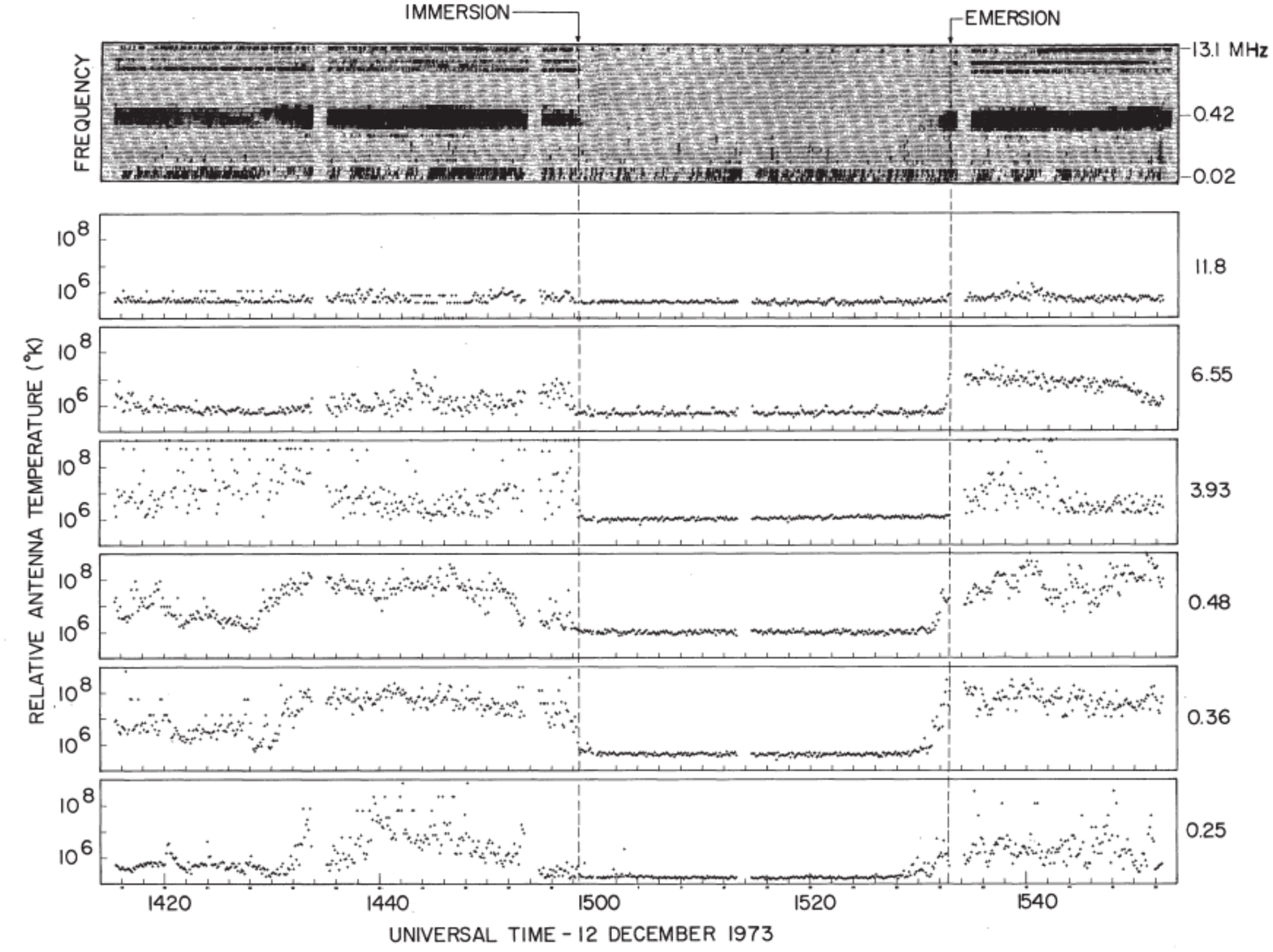}
\caption[]{Shielding of terrestrial radio interference by the moon, as observed by the RAE-2 satellite. Even at a distance of 300,000 km from the Earth, the additional shielding by the Moon provides 1-3 orders of magnitude (10-30 dB) of additional interference suppression. Taken from \protect\citet[\copyright\ ESO]{AKNea75}\label{fig:moonquiet}}
\end{figure}

So far, there have only been two space missions whose primary purpose was low-frequency radio astronomy: the Radio Astronomy Explorers (RAE) 1 and 2. It was a surprising finding of RAE-1 that the Earth itself is a strong emitter of low-frequency bursts, the so-called Auroral Kilometric Radiation (AKR) generated by solar-wind interactions with the Earth's magnetic field. This emission is so strong that RAE-2 was placed in a lunar instead of the originally planned terrestrial orbit to provide shielding from this natural terrestrial interference, see Fig.\ref{fig:moonquiet}. The structure of the Galaxy's emission was only seen by RAE-2 at very low spatial resolution (with beam sizes between 37$^{\circ}~\times~61^{\circ}$ and 220$^{\circ}~\times~160^{\circ}$) and fairly low signal-to-noise ratio. Due to the very limited angular resolution and the large power of AKR, it proved impossible to image any discrete sources directly. 

Placing a low-frequency radio antenna array in an eternally dark crater or behind a mountain such as Malapert (see also Section~\ref{sec:21cm}) at the lunar South Pole would potentially provide excellent shielding from man-made radio interference (RFI), absence of Earth's ionospheric distortions, and long-term antenna gain stability. That would allows for sensitive and high resolution low-frequency radio astronomy observations addressing a wealth of science cases and essentially opening up the last virtually unexplored frequency regime for astronomy. However, as have been discussed in the proceeding sections: not even the Moon is entirely free from distortions. Therefore, during the course of the routine observations of the lunar dust and plasma environment that are proposed to be performed by L-DEPP, a knowledge base will be built up, which allow for the first experimental assessment of the Moon as a potential site for such astronomical ultra-long wavelength radio observations.

Despite these environmental complications, the potential for a single tripole antenna on the moon is still large and below we summarize the \emph{astronomy} science cases that we have identified such an antenna placed on a relatively stable location on the south pole of the moon can address \citep[see also][]{JF2009}. 

In addition to the science cases described below, the LRX will open up the last virtually unexplored frequency domain. This will undoubtedly lead to the discovery of new and unexpected phenomena, just like any previous new instrument that gave access to unexplored  regions of the electromagnetic spectrum have done in the past.

\subsection{The Radio background}
\label{bkg}

The charging of the lunar surface, the dusty plasma and electromagnetic (radio) radiation may also affect future lunar instruments and even radio communications on the surface. For example, part of a future lunar exploration program will undoubtedly be the erection of some larger technical and scientific experiments. ESA's large Exploration Architecture Review mentions as a key example for science from the Moon the construction of a low-frequency radio telescope. The reasons are that such telescopes are easy and light-weight to build but need to be outside Earth's ionosphere, require shielding from RFI, and a large stable ground all of which is naturally provided by the Moon. The interesting frequency ranges for such an array are $1-10$ MHz, where low-frequency radiation is completely reflected by the ionosphere of the Earth and $10-100$ MHz, where the incoming phase of the radiation is distorted by the ionosphere (similar to seeing for optical telescopes). As preparation, one requires in-depth understanding of the dielectric properties of the Moon and of the spectral and temporal radio background.

There is a spectral component to the radio background which simply comes from the variations of the lunar exosphere. Below 1 MHz the AKR of the Earth dominates the RFI, and above 1 MHz solar Type II and Type III bursts, associated with solar flares and CMEs, as well as the Jovian decametric radiation, would cause intermittent disturbances. Solar and AKR events may also be potential drivers of lunar dust activity.

\subsection{Cosmic Ray Impacts}
\label{cosmic rays}

The temporal radio background on the lunar surface has never been studied so far. What makes the Moon so different is again its large dusty surface, exposed to free space. For example, Ultrahigh Energy Cosmic Rays (UHECR) and neutrinos will impact the Moon surface directly and produce radio emission, e.g. through the so-called Askaryan effect \citep[see e.g.][]{scholten06,staal2007,JF2009}. Ultrahigh energy cosmic rays in particular are rare, with an estimated flux of less than one neutrino per km$^{2}$ and year over 2$\pi$ steradian, calling for a huge detector. Already Askaryan himself suggested using the Moon as a detector in his original paper from 1961. On Earth, the Antarctic ice sheet and the Mediterranean Sea have been used but although the Askaryan effect has been confirmed in the lab (by SLAC), it still awaits confirmation in nature. Nonetheless, it has already been used by a number of experiments, including two NASA balloon missions (ANITA light and ANITA) and experiments where radio telescopes observe the Moon from Earth. Estimates \citet{JF2009} suggest that this can provide quite a high local background, but potentially also provides an interesting diagnostic tool to identify nearby cosmic ray impacts.

\subsection{Cosmology with the 21 cm Hydrogen line}
\label{sec:21cm}

A fundamental question of current cosmological research is on the nature of structure formation in the Universe \citep[for reviews, see][]{CiardiFerrara05,fur2009,prit2010}: how is the observed distribution of visible matter created from the initial conditions just after the big bang, when matter and radiation were distributed extremely smoothly, with density variations of just one part in 100,000? The CMB radiation was emitted at z$\approx1200$, about 400,000 years after the Big Bang, when the Universe had cooled off sufficiently for electrons and protons to recombine into neutral hydrogen atoms, allowing the background radiation to move freely without being scattered by the electrons and protons. However, although there were plenty of photons, these were scattered randomly, there were no sources of light in the Universe yet: the hydrogen and helium that were created in the big bang first had to cool in order to be able to clump together and form stars and galaxies. Hence, this era is called the Dark Ages, in the redshift range z = $30-1000$ \citep{Rees99}.

Things only changed after the first stars, galaxies and active black holes had formed and emitted enough UV and X ray photons to reionize the neutral hydrogen. Compared to the epoch before Recombination, the Universe had expanded so much that the plasma was sufficiently diluted to let the radiation pass freely. The time when this happened is called the Epoch of Reionization (EoR) and is believed to have occurred around z=11, about 400 million years after the Big Bang, though it is at present not known whether the reionization happened more or less instantaneously, similar to a global phase transition, or was more or less spread out in time, depending on local conditions.

Throughout all the Dark Ages and the EoR hydrogen played a major role, emitting or absorbing the well-known 21 cm (1.4 GHz) line due to the spin flip of the electron. Today, this emission is redshifted by a factor $10-1000$ and lies in the $1.4-140$ MHz frequency range. When the hydrogen spin temperature is not coupled perfectly to the radiation temperature of the cosmic background radiation, but changed by other couplings with the surrounding matter and radiation, it can be seen against the cosmic background radiation in absorption or emission, depending on whether the spin temperature is lower or higher than the background radiation temperature.

\subsubsection{Expected strength of the 21 cm line}

Observing the cosmological 21 cm emission is the only known method to directly probe the Dark Ages and the EoR. Therefore it is subject of a rapidly developing literature, which is reviewed by \citet{FOB06}. Experimental verifications have been performed by the EDGES experiment \citep{br2010} and we are aware of at least one space mission proposal, the Dark Ages Radio Explorer (DARE) \citep{burns2011,burns2012} that intends to measure the red shifted 21 cm spectrum using an orbiter around the Moon, with the Moon acting as a shield for RFI from Earth. 

\begin{figure}
\centering
\includegraphics[width=\columnwidth]{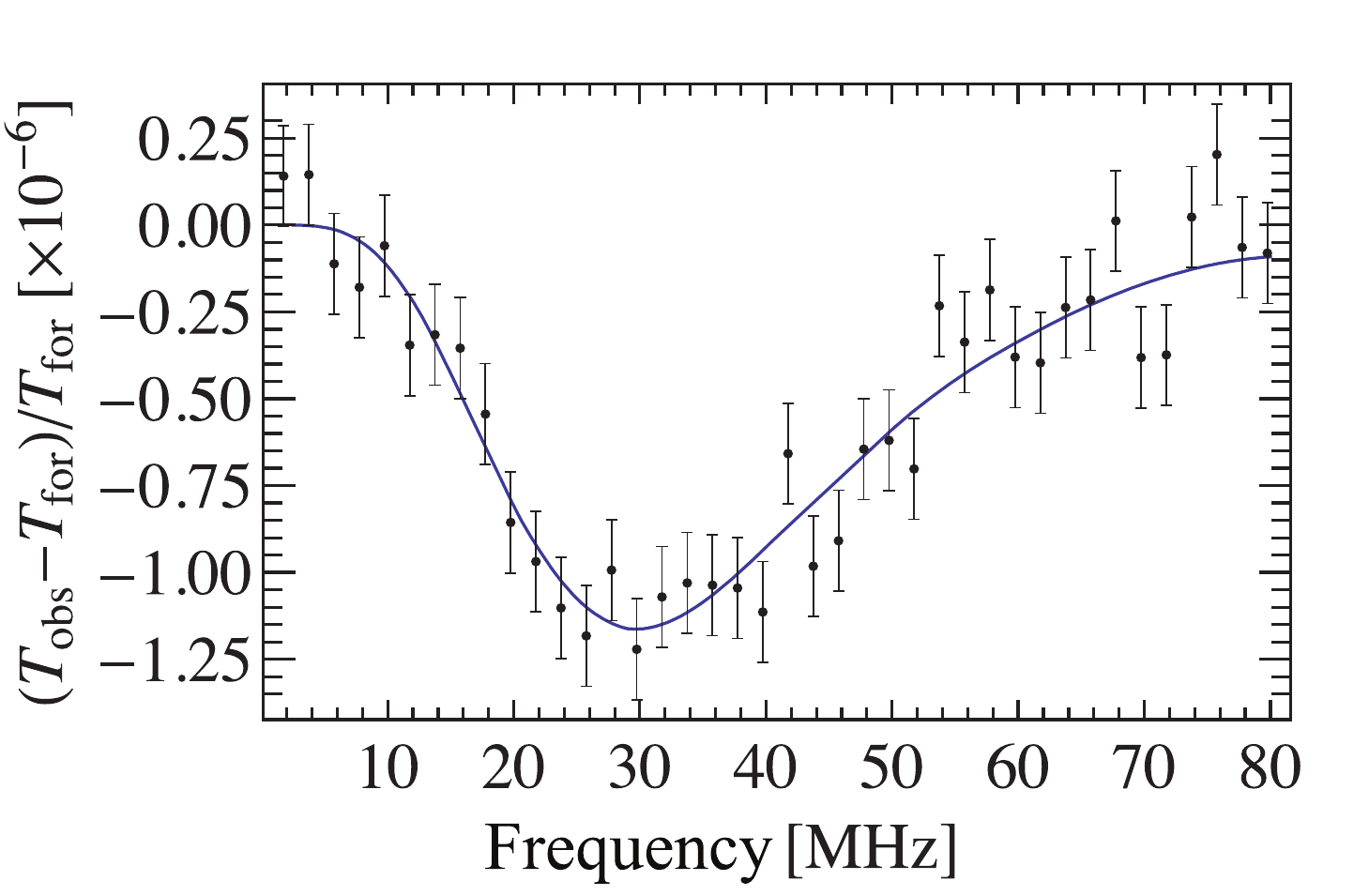}
\caption{A simulated global 21-cm HI signal from the dark ages after 1 year of integration time with a single sky-noise limited dipole as a function of frequency \citep[based on the calculations by][]{CS07}. Here a fixed bandwidth of 1.5 MHz is used. We assume that the observed brightness temperature, $T_{\mathrm{obs}}$, is just the sum of the dark ages signal and a foreground signal, $T_{\mathrm{for}}$, which is a perfect power law. The signal at 30\,MHz originates from $z=46$.\label{fig:darkages}}
\end{figure}

A single antenna with an instantaneous frequency range $10-200$ MHz, in an eternally dark lunar crater or behind a large mountain range at the pole, kept at a stable temperature, and shielded from any solar and terrestrial radiation, would possibly offer one of the very best set-ups for such an experiment. That should reduce most instrumental and RFI effects to the bare minimum and only leave the cosmic foreground as the main issue. The simulated 21 cm line detection with Gaussian noise in Fig.~\ref{fig:darkages} shows that the signal is only $10^{-6}$ of the foreground signal. Moreover, detection has to rely on the foreground to be a perfect power law. Hence, the foreground subtraction is the biggest uncertainty together with the need for an almost perfect band-pass calibration of the antenna. It is likely that the actual foreground signal will deviate from a simple power-law. This makes detection almost impossible if there are spectral foreground fluctuations of the same magnitude as the dark ages signal. Nonetheless, further studies need to show on which level spectral foreground fluctuations are actually present. As in the global EoR case, the global dark-ages 21 cm signal is isotropic, so that a single (well-calibrated) dipole suffices to carry out this measurement and detect the signal in frequency space. The concept recommendation for the LRX proposes a tripole antenna consisting of three individual dipoles. This would make the tripole also three times more sensitive, and hence would require three times shorter integration time to reach the same sensitivity.

\subsubsection{expected RFI levels and attenuation at the Lunar South Pole}
   
With the expected signal being extremely weak compared to the background noise, one of the major issues is what is the expected strength of the RFI noise component and how much attenuation will the Lunar South Pole location offer? Here we present an initial discussion on this issue based on some preliminary analysis that we have performed; a more detailed analysis of the expected RFI level and simulations of the attenuation at the ELL landing site taking into account all local environmental conditions will be presented separately.

The nominal RFI level (without attenuation by the Lunar environment) at the moon is the combination of the component of all the man-made RFI that penetrates the Earth's ionosphere and the natural RFI noise (e.g. AKR). This nominal level would be very much dependent on the location, as one can imagine that a strong radio transmitter beamed in the line of sight would have a large contribution. Hence, attempting to estimate this nominal RFI by using Earth-based instruments would not appear to be correct. Instead, a more reliable estimate is provided by space-based missions such as RAE-2 \citep{NB78}, WIND/WAVES \citep{bougeret95,kaiser96}, FORTE \citep{burr05}. Table~\ref{table:rfi} presents the RFI levels as measured by these instruments and from this table it is clear that comparing these results is difficult as both their locations and frequency band of observation were different. In fact, most surprisingly the only two missions that observed at similar frequencies (10 MHz), RAE-2 and WIND/WAVES, show similar RFI levels (around 40 dB) while RAE-2 is about a factor 2 further from Earth. This supports the idea that the nominal RFI level depends strongly on the location it is measured. \emph{As the RAE-2 mission was in orbit around the moon, we assume a level of 40 dB above the Galactic Background level as the nominal RFI at the Lunar South Pole.}

\begin{table}[ht]
\caption{Overview of RFI measurements from space-based missions. The RFI is the level in dB measure above (positive) or below (negative) the Galactic Background level. Note that for WIND/WAVES the RFI level measured in the two bands is very similar, i.e. dB levels up to 40 \citep[see][]{bougeret95}. }
\scriptsize
\begin{tabular} { l l c c c } %
\hline\hline 
Mission & Location & Freq. (MHz) & RFI (dB)  & Date \\  
\hline
\hline
REA-2 & Moon & $1-10$ & $30-40$ & 1970s \\
 & Orbit & & &\\
\hline
WIND/& 200.000 km& 6.125,& $<40$ & 1994 \\
WAVES & Earth Orbit &10.325 & &\\
\hline
FORTE & 800 km & 38 & $40$ & 1997 \\
 & Earth Orbit & & &\\
\hline
FORTE & 800 km & 130 & $-10$ & 1997 \\
 & Earth Orbit & & &\\
\hline
\end{tabular}
\label{table:rfi}
\end{table}  

Attenuation of the nominal RFI signal can originate from the Lunar ionosphere, rocks and mountains on the moon and of course the moon itself. Recently \cite{burns2012} suggested that attenuation in the order of 80 or 90 dB is expected to be required for detecting the 21-cm above the RFI signal, and that such strong attenuation can only be achieved at some far-side locations. \cite{taka03} has already shown that the Malapert Mountain area is a promising area for low frequency radio astronomy as attenuation up to $40-60$ dB can be reach for certain locations. In that work, \cite{taka03} also simulate the diffraction of the RFI signal and use a simple sinusoidal topology to show how radio waves at 0.5 and 1.0 MHz are diffracted around Malapert Mountain. From these simulations it is clear that the resulting attenuation depends strongly on the location but that locations can be found that are associated with attenuation ranging from $\sim20$ to $\sim40$ dB.Note that the results from \cite{taka03} do not cover the large frequency range covered by the LRX and moreover, are outside the frequency band where the 21 cm line is expected to peak. However, as higher attenuation is expected at larger frequencies the RFI level could be further reduced. Our own preliminary simulations at 10 MHz and 30 MHz confirm these results, and moreover indicate that a mountain like Malapert absorbs a significant fraction of the incident radiation causing "RFI-shadow" area's behind a mountain like Malapert.

Concluding, we tentatively find that the Lunar South Pole offers locations with significant shielding of the RFI noise component. However, given the fact that the exact ELL landing site still has to be selected, the uncertainty in the nominal RFI level, the strong dependence of the attenuation on the local ionospheric and geophysical conditions and the influence of the diffraction of the RFI signal, we urge for the need of more detailed (2D and 3D) simulations. Ultimately in-situ measurements at the Lunar South Pole will be required to accurately determine the RFI level, the radio background spectrum and local attenuation which will pave the way for detailed study of the early universe through the 21 cm line emission. 

\subsection{Planetary Science}
\label{planets}

The strong magnetic field of the Earth and of the 4 solar system giant planets carves a cavity in the solar wind, the magnetosphere, where electrons are accelerated to keV-MeV energies by various processes, and give birth to intense non-thermal low frequency radio emissions near/above the magnetic poles (so-called auroral regions). The radio emission is produced at the local electron cyclotron frequency $f_{ce} = (eB)/(2 \pi me)$ with B the magnetic field intensity at the source, or $f_{ce}$(MHz) = 2.8 B (Gauss), by a resonant mechanism that allows electrons with keV energy to transfer a large amount of their energy (up to a few $\%$) to electromagnetic (radio) waves. This mechanism, called the Cyclotron Maser instability (CMI) is by far the most efficient low frequency radio generation mechanism, and it operates at all the radio planets in our solar system \citep{Zarka1998}. 

Figure 1 of [Zarka, 1998] displays normalized average spectra of all solar system radio planets (an update on Jupiter is in \citep{Zarka2004}). Spectral flux densities at Moon surface are discussed in more details in Zarka et al. [this issue]. Average values reach a few $\times10^{5}$ Jy for Jupiter, $\sim2\times10^{4}$ Jy for Saturn, and about $10^{2}$ Jy for Uranus and Neptune. Peak spectra are about one order of magnitude higher. Due to its intense magnetic field (up to 14 G at the surface), Jupiter's emission reaches 40 MHz, while that from other planets lies below 1.5 MHz. All emissions are broadband, because their sources stretch along high-latitude magnetic field lines from the planetary surface to far above it, and they are detected with a complex morphology in the time-frequency plane because each magnetosphere may contain several sources of accelerated electrons (e.g. auroral or induced by the Io-magnetosphere interaction at Jupiter) and because the emission beaming due to CMI is very anisotropic, peaking at large angle from the local magnetic field. The typical dynamic spectrum of Jupiter's magnetospheric radio components below 16 MHz, shown in Figure 6a of \cite{Zarka2004}, illustrates this complex morphology.

Jupiter's radio emissions are intense, circularly or elliptically polarized (depending on their magnetic hemisphere of origin), and instantaneously point-like at a given frequency. The activity is quasi-permanent in the hecto-kilometer range, and reasonably predictable at decameter wavelengths \citep{GZL1989}. The full Jovian radio spectrum will be detectable by the LRX experiment with integrations of order of $\sim1$ sec $\times~10$ kHz. It will provide a good probe of the Lunar ionosphere, as well as a good calibrator for the received flux by comparison with simultaneous ground-based observations above $\sim10$ MHz. Saturn's radio emissions should be detectable with slightly broader integrations (a few sec $\times$ a few 10's kHz) and Uranus and Neptune might be detectable with $\sim100$'s sec $\times 500$ kHz integrations [Zarka et al., this issue], provided that the antenna's preamplifier current noise is low enough.

For all these giant planets, regular observations from a quasi-fixed vantage point should allow to study the time variability of the radio emissions, from short pulses to planetary rotation period, modulations satellites and by the solar wind, and up to seasonal and solar cycle effects. In particular, accurate determination of the planetary rotation periods has become a subject of extremely high interest with the discovery of the still unexplained dual and variable radio period of Saturn \citep{lamy2011}. Modulations of radio emissions due to natural satellites and to the solar wind strength will allow to address magnetospheric dynamics, solar wind-magnetosphere coupling (substorms), electrodynamic coupling of the magnetosphere with embedded moons. They will also permit to monitor the solar wind from 1 to 30 AU (with the magnetospheric radio emission intensity as a proxy). At Jupiter, polarization of the decameter emission provides a way to probe the Io plasma torus via Faraday rotation measurements. If Uranus and/or Neptune's radio emissions are detected, it will be for the first time since the Voyager 2 mission in the late 1980's. LRX will thus offer a unique opportunity in decades to further study the very peculiar dynamics of their magnetospheres of origin, with tilted rotation and magnetic axes, leading to helicoidal plasma dynamics at Uranus \citep{Arridge2011} and periodic pole-on magnetospheric configuration at Neptune \citep{Chris2011}.

Sporadic radio emissions associated to atmospheric lightning discharges have been detected from Saturn and Uranus by the Voyager and Cassini spacecraft. Typical burst durations are $\sim30-500$ msec but their long-term occurrence is very variable \citep{Zarka2008}. With 3 dipoles, only the strongest flashes from Saturn will be detectable [Zarka et al., this issue], prolongating the monitoring that Cassini will perform until the end of the mission in 2017, which permits to address questions such as meteorology and seasonal variations (related to atmospheric dynamics), and Saturn's ionospheric density (via low-frequency cutoff of lightning radio signals). With only a few dipoles, exoplanetary magnetospheric radio emissions are beyond reach \citep[][ Zarka et al., this issue]{Zarka2007}, but LRX measurements will set the context for future exoplanet low frequency radio searches with a lunar array.

Even with modest signal-to-noise ratio, correlation of the signals receives by the 3 orthogonal dipoles on LRX will allow to determine the wave vector and full polarization of incoming radio waves \citep{Cecconi2010}. Whenever a single source dominates a instantaneous measurement (this should be often the case because planetary radio emissions are sporadic) its direction will be determined with a few degrees accuracy. This will allow us to distinguish planetary radio emissions from each other.

LRX planetary observations will complement higher frequency ($\ge10-30$ MHz) ground-based observations with large ground-based instruments such as LOFAR, UTR2 (Kharkov), or the Nancay decameter array \citep{GZG2011}, and provide context for in-situ spacecraft measurements. Cross-calibration of intensity and polarization, stereoscopic observations, follow-up high sensitivity targeted observations, and even very long baseline interferometric correlations are all to be considered. Radio observations will also be complemented by observations at other wavelengths (e.g. with UV images from Hubble Space Telescope \citep{Lamy2012} or its successor).

\subsection{Solar Physics}
\label{sun}

The Sun, our closest star, fundamentally affects the Earth's ecospace and our daily life. Our Sun is a very strong radio source: Superimposed on the thermal emissions of the quiet sun are the intense radio bursts that are associated with solar flares and CME's, clouds of ionized plasma ejected into interplanetary space. Three main types of radio bursts are observed from the Sun, particularly in its active state, both relate to flares and CME's. Type II bursts have a frequency drift with time at rates consistent with the speed of the shock through the solar corona and interplanetary medium ($\sim1000-2000$ km/s). Type III bursts are emitted by mildly relativistic ($\sim0.1 - 0.3$ c) electron beams propagating through the corona and interplanetary space that excite plasma waves at the local plasma frequency. Their frequency drift rate is much higher than that of Type II bursts. Type IV bursts are emitted by energetic electrons in the coronal magnetic field structure (such as coronal loops). Both the Type II and III bursts can be imaged by a single tripole on the Lunar South Pole in the $<1$ to $30$ MHz range \cite[but see also the ROLSS concept for a radio antenna on the lunar near side][]{lazio2011}.

The density model of the heliosphere \citep{Mann99} directly relates the radio source location (in solar radii) to the emission frequency: higher frequency radio emission originates closer to the surface of the sun, while lower frequency emission originates further out. Dynamic radio spectrograms and coronal height-time diagrams thus provide detailed information of the movement of plasma through the solar corona and out into interplanetary space. Through the magnetospheric-ionospheric coupling such events have a direct influence on the lunar ionosphere. However, the physical mechanisms that govern them are not fully understood. By providing dynamic spectra and detailed imaging of the solar radio emissions, one will be able to monitor and model the plasma instabilities in the solar corona and wave-particle interactions in the activity centers of the Sun.

A low frequency radio antenna on the south pole of the Moon offers great opportunities for radio studies of the solar wind and the heliosphere and allows for observations much further out from the solar surface than possible from the ground, where the ionosphere confines the field of view to within a few solar radii. One would potentially dynamically image the evolution of CME structures (or radio bursts and flares) as they propagate out into interplanetary space and potentially impact the creation of dust on the Moon or alter the conditions of the lunar ionosphere. Observing at frequencies down to 0.1 MHz one can potentially follow and image CME's out to 2 AU, particularly if they are on a collision course with the telescope itself. The combined in-situ and remote solar observations will allow us to track in detail the effect of solar activity on the lunar environment directly. A south pole location would allow for observing the effects related to a single CME, flare or radio burst, of the solar wind on the creation of ions in both the dark and sunlit side of the Moon, as well as the long-term effects on the plasma on the dark side.

In addition, a single radio antenna on the Moon will also provide an ideal low-noise facility for Interstellar Plasma Scintillation (IPS) observations in its own right, with the additional virtue of providing heliospheric calibration for all-sky astronomical measurements. Even if confusion limited and with limited instantaneous imaging ability, low frequency lunar radio experiments conducted simultaneously with terrestrial LOFAR observations will provide a series of proof-of-concept low- frequency scintillation observations in preparation for the design of more extensive future arrays. Complementary terrestrial LOFAR observations will provide the heliospheric context for lower frequency imaging with a lunar based antenna.

\section{Preliminary LRX design concept}
\label{design}

\begin{figure}
\centering
\includegraphics[width=\columnwidth]{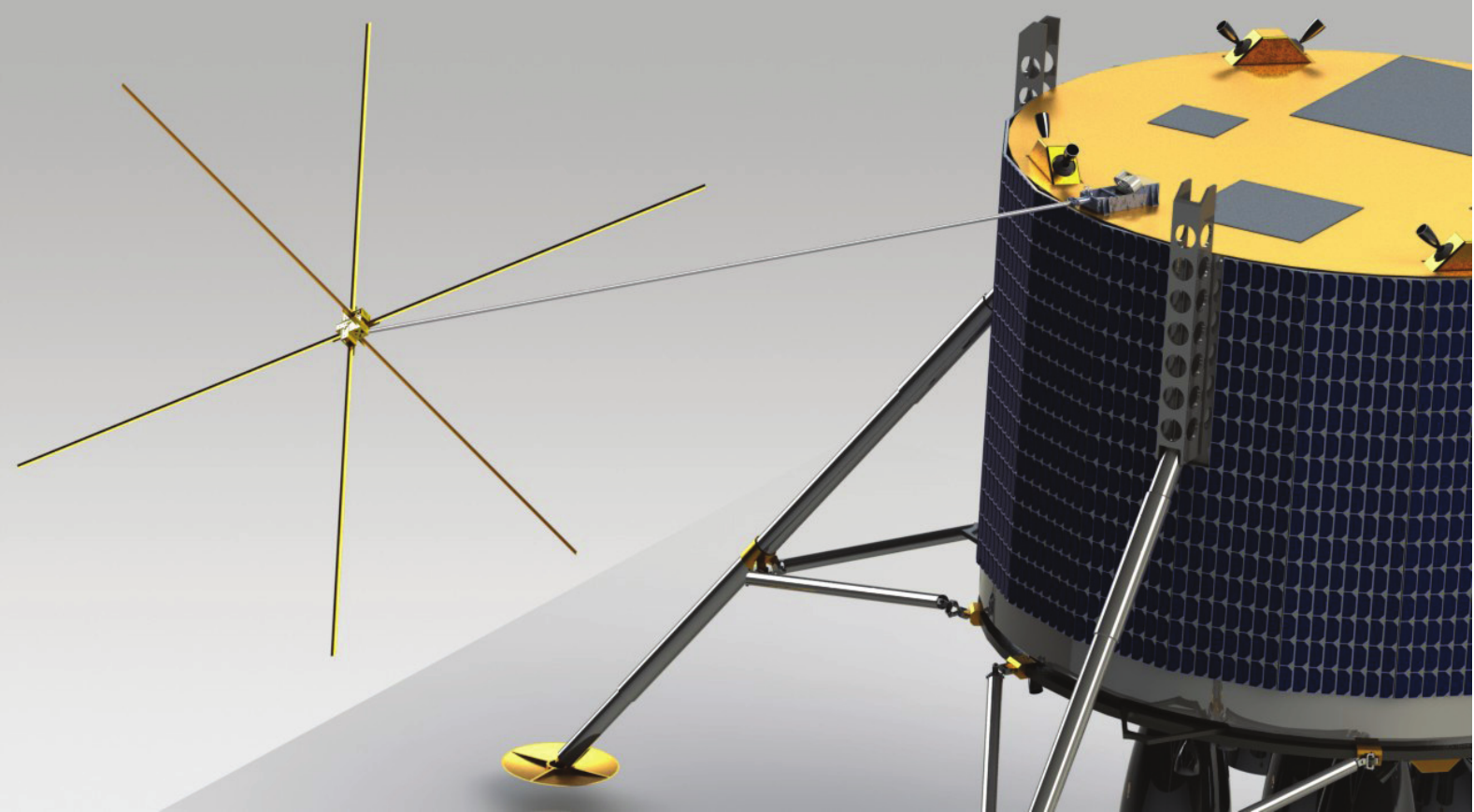}
\caption[]{The LRX on the Lunar Lander. \label{fig:lrxlander}}
\end{figure}

A number of different antenna configurations for the LRX have been studied, but based on arguments related to the performance (i.e. resulting beam pattern, radiation efficiency, polarization, SNR and DOA capabilities in the frequency range of interest) and the deployment a design was selected with an active tripole antenna consisting of 3 individual dipoles, each 2.5 m in length. The tripole shall be mounted on a boom so that its arms are symmetrically aligned in the xyz directions. Due to the symmetry, the radiation pattern would be omnidirectional and the algorithm for direction of arrival signals requires less memory. The choice of orientation depends on the resulting beam pattern, the interaction with the lander and deployment mechanisms and shall be studied in more detail. However, to make a symmetrical configuration, the angle between each dipole and the horizontal plane could be $35.3^{\circ}$. Note that the boom for mounting the antenna can be conductive and similar (in size and material) to the one used for mounting the Langmuir probes (which are also part of the L-DEPP package). In Fig.~\ref{fig:lrxlander} we show the tripole antenna mounted on a boom which is mounted on the upper deck of the Lunar Lander.

The antennas will be made out of Beryllium alloys, which have good strength and conductivity. The current choice is between Aluminum-Beryllium (AlBeMet), which is light, and Copper-Beryllium, which has good strength and has been already used for space missions (e.g. Ulysses Radio/Plasma antennas). With the thickness of 0.1 mm for a U- shape tape the latter is the preferred option, in which case the mass of each arm (1.25 m) would be around 20 grams. Deployment for the tripole antenna can be such that each arm of the tripole (in total 6 arms) is placed in a tape housing with dimensions of about $3\times3\times2$ cm. Each housing will be fitted with a thermal knife-type releasing mechanism. The housing is proposed to be made of Ultem 1000, a light, non-conductive material with enough strength. The mass of each antenna tape with its housing and LNA is less than 50 grams. Prototypes of such a tape-measure deployment mechanism have been developed at the Radboud University Nijmegen, and will be developed in more detail. Another option for deployment for the tripole antenna is to simply fold the individual arms in the direction of the boom during flight and unfold them once on the surface. Both options will be studied in more detail.
 
\begin{figure}
\centering
\includegraphics[width=\columnwidth]{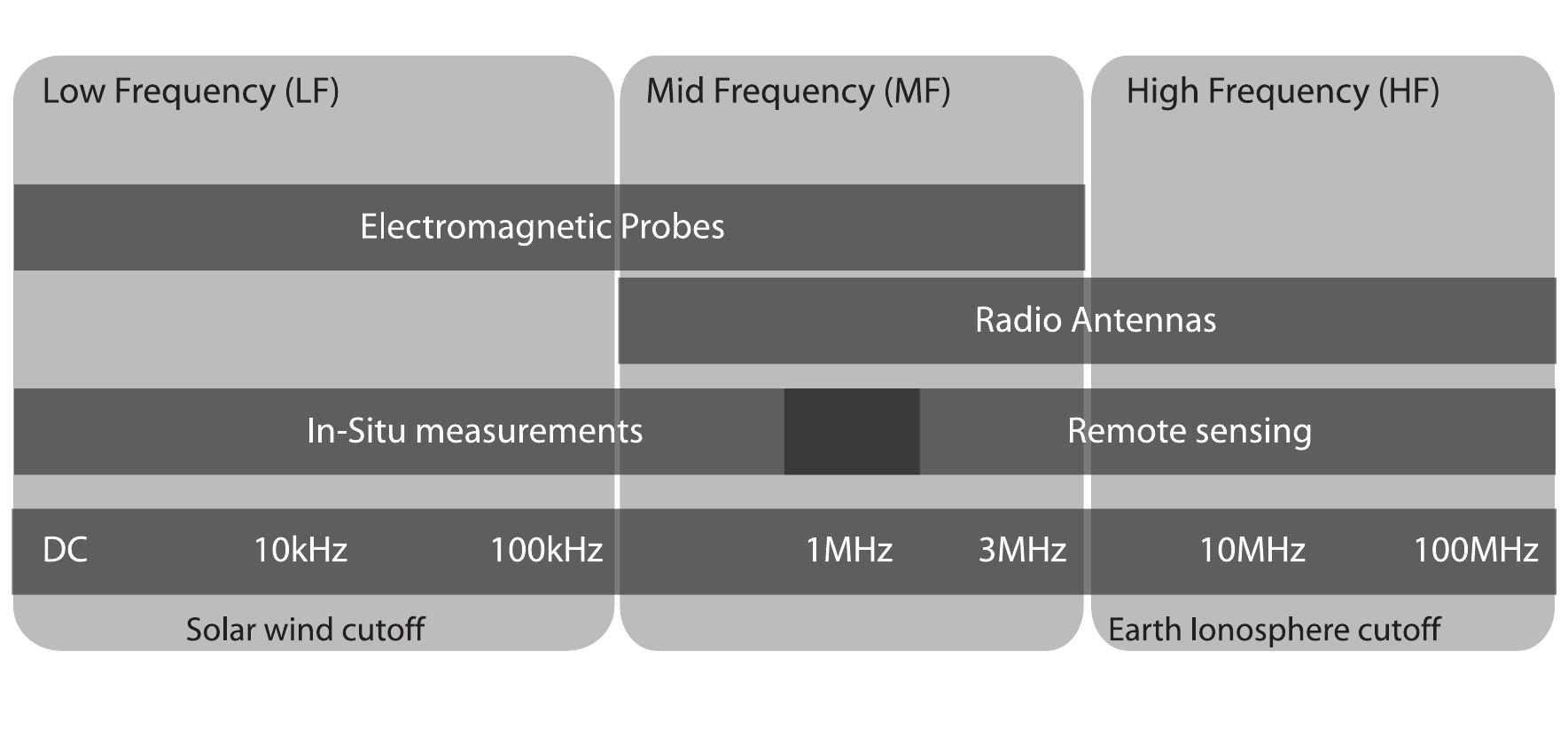}
\caption[]{The frequencies bands defined for the L-DEPP radio and Langmuir Probe instruments. There are three bands defined, Low, Mid and High Frequency; the LRX will only operate in the Mid and High Frequency bands. The vertical dark grey bands indicate (from the top down) the type of instrument that operates in the corresponding frequency range, the type of measurements that can be performed and finally the frequency range is  shown at the bottom. \label{fig:freq}}
\end{figure}

The LRX tripole antenna will have its own LNA's, filter and amplifier in the tripole center. The onboard digital processing process is then performed by a dedicated PSD board: first an ADC digitizes data, then this digital data goes to the DPU which includes an FPGA module and a microprocessor for triggering, processing and analysis. The data acquisition system is required to be able to make the following processing steps:
\begin{itemize}
\item{Fast Fourier transform of the wave form to obtain spectra}
\item{Integration and inspection of spectra}
\item{Buffering of wave forms for offline-processing and pulse localization}
\item{Detecting and triggering on radio burst}
\end{itemize}
Current design requirements are for a 3-channel system with up to 200 MHz sampling rate. These requirements appear feasible, but further study is likely to be required. The low and high frequency regime will be processed separately due to the additional noise patterns present at frequencies below 1 MHz. The following bands are suggested: 
\begin{itemize}
\item{Mid Frequency or MF band: $100$ kHz$-3$ MHz}
\item{High Frequency or HF band: $3$ MHz$-100$ MHz}
\end{itemize}
The main advantage of this approach is that it is not required to process the whole broad frequency band in one go, which would introduce additional noise. The choice of frequencies is made such that there it matches up with the frequencies for the Langmuir Proves that are also part of the L-DEPP package, see Fig.\ref{fig:freq} for an overview.

Finally, in Table~\ref{table:req} we summarize the technical requirements as they follow from the science cases as presented in the previous sections.The numbers presented in this table are calculated taking into account a nominal mission life time of one year, the memory storage and data transfer limitations on the ELL.Depending on the sensitivity the integration time required to detect radio sources may vary from seconds (e.g. noise background) to the mission life time (e.g. 21 cm Dark Ages signal). Activities are planned for optimizing and testing both the front- and the back-end of the LRX, addressing key issues such as the extreme temperature conditions (ranging from 30 K to 300 K) the electronics (for instance the LNA) are exposed to,the required sensitivity and spectral response for a single active tripole antenna covering such a broad frequency range, calibration of the antenna taking into account the effects of the ELL and the Lunar surface, implementation of RFI mitigation techniques \citep[e.g.][]{fridman01} and the removal of foreground objects \citep[for detailed discussions see][]{JF2009,harker2009,chap2012,harker2012} for optimizing the receiver performance for detecting the weak 21 cm line. Here we will also use the results from ongoing (data and instrument) calibration and testing of the LOFAR antennas.

\begin{table}[ht]
\caption{Summary of LRX System Requirements.}
\centering
\scriptsize
\begin{tabular} { c c c c c }
\hline\hline
Experiment & Frequency  & Temp.  & FFT& Snapshots\\
	&	range & res.& res.& per day\\
\hline
\hline
Lunar & 100 Hz& 300 s & 1 KHz & 100 \\
Plasma & $-50$ kHz & & &\\
\hline
Lunar &  5 kHz & 300 s &1 KHz & 50 \\
Exosphere & $-3$ MHz & & &\\
\hline
Solar and &    1MHz &600 s & 200 KHz & 50\\
Planetary& $-40$ MHz & &\\
Science & & & &\\
\hline
Astronomy& 10 KHz& 1200 s & 200 KHz & 50 \\
and &$-100$ MHz & & &\\
Cosmology & & & & \\
\hline
\end{tabular}
\label{table:req}
\end{table}

\section{Conclusions}

The European Lunar Lander is envisaged to land on the South Pole of the moon in 2019 and among a multitude of instruments a low frequency, active tripole antenna (LRX) is considered to be part of the payload. The carefully selected South Pole location not only provides a unique view of both the sun-lit and the dark side of the moon simultaneously, allowing for a detailed mapping of the lunar ionosphere, but also potentially provides the shielding from the man-made RFI, absence of ionospheric distortions, and high temperature and antenna gain stability that allows detection of the 21 cm wave emission from pristine hydrogen formed after the big bang and into the period where the first stars formed. The LRX antenna opens up the last virtually unexplored frequency regime below $\sim30$ MHz for radio astronomy, allowing for study of  auroral radiation in the atmosphere of Jupiter and Saturn, detection of high-energy cosmic rays, and more detailed study of the solar CME to distances closer to earth (space-weather prediction). More importantly, the LRX is a pathfinder mission and will determine  by in-situ measurements the Lunar radio background in unprecedented detail, paving the way for a future large low-frequency radio array on the moon.

\section{Acknowledgements}
The authors would like to thank Prof. Jan Bergman of the Swedish Institute of Space Physics, Prof. Ralph Srama of the University of Stuttgart and Gerrit Hausmann, Dr. Juergen Burfeindt and Hans-Guenter Bernhardt of Kayser-Threde GmbH Germany, Prof. dr. L.V.E. Koopmans of the Kapteyn Astronomical Institute, Prof. O. Scholten of the "Kernfysisch Versneller Instituut" and Dr. Stijn Buitink and Thomas Bronzwaer of the Radboud University Nijmegen for all their useful discussions and contributions. In addition the authors thank the referees for their useful comments and suggestions


\begin{thebibliography}{57}
\expandafter\ifx\csname natexlab\endcsname\relax\def\natexlab#1{#1}\fi
\expandafter\ifx\csname url\endcsname\relax
  \def\url#1{\texttt{#1}}\fi
\expandafter\ifx\csname urlprefix\endcsname\relax\def\urlprefix{URL }\fi

\bibitem[{{Alexander} et~al.(1975){Alexander}, {Kaiser}, {Novaco}, {Grena}, and
  {Weber}}]{AKNea75}
{Alexander}, J.~K., {Kaiser}, M.~L., {Novaco}, J.~C., {Grena}, F.~R., {Weber},
  R.~R., May 1975. {Scientific instrumentation of the
  Radio-Astronomy-Explorer-2 satellite}. Astronomy and Astrophysics 40,
  365--371.

\bibitem[{{Allan} and {Poulter}(1992)}]{allan92}
{Allan}, W., {Poulter}, E.~M., May 1992. {ULF waves-their relationship to the
  structure of the Earth's magnetosphere}. Reports on Progress in Physics 55,
  533--598.

\bibitem[{{Arridge} et~al.(2011){Arridge}, {Agnor}, {Andr{\'e}}, {Baines},
  {Fletcher}, {Gautier}, {Hofstadter}, {Jones}, {Lamy}, {Langevin}, {Mousis},
  {Nettelmann}, {Russell}, {Stallard}, {Tiscareno}, {Tobie}, {Bacon},
  {Chaloner}, {Guest}, {Kemble}, {Peacocke}, {Achilleos}, {Andert}, {Banfield},
  {Barabash}, {Barthelemy}, {Bertucci}, {Brandt}, {Cecconi}, {Chakrabarti},
  {Cheng}, {Christensen}, {Christou}, {Coates}, {Collinson}, {Cooper},
  {Courtin}, {Dougherty}, {Ebert}, {Entradas}, {Fazakerley}, {Fortney},
  {Galand}, {Gustin}, {Hedman}, {Helled}, {Henri}, {Hess}, {Holme},
  {Karatekin}, {Krupp}, {Leisner}, {Martin-Torres}, {Masters}, {Melin},
  {Miller}, {M{\"u}ller-Wodarg}, {Noyelles}, {Paranicas}, {de Pater},
  {P{\"a}tzold}, {Prang{\'e}}, {Qu{\'e}merais}, {Roussos}, {Rymer},
  {S{\'a}nchez-Lavega}, {Saur}, {Sayanagi}, {Schenk}, {Schubert}, {Sergis},
  {Sohl}, {Sittler}, {Teanby}, {Tellmann}, {Turtle}, {Vinatier}, {Wahlund}, and
  {Zarka}}]{Arridge2011}
{Arridge}, C.~S., {Agnor}, C.~B., {Andr{\'e}}, N., {Baines}, K.~H., {Fletcher},
  L.~N., {Gautier}, D., {Hofstadter}, M.~D., {Jones}, G.~H., {Lamy}, L.,
  {Langevin}, Y., {Mousis}, O., {Nettelmann}, N., {Russell}, C.~T., {Stallard},
  T., {Tiscareno}, M.~S., {Tobie}, G., {Bacon}, A., {Chaloner}, C., {Guest},
  M., {Kemble}, S., {Peacocke}, L., {Achilleos}, N., {Andert}, T.~P.,
  {Banfield}, D., {Barabash}, S., {Barthelemy}, M., {Bertucci}, C., {Brandt},
  P., {Cecconi}, B., {Chakrabarti}, S., {Cheng}, A.~F., {Christensen}, U.,
  {Christou}, A., {Coates}, A.~J., {Collinson}, G., {Cooper}, J.~F., {Courtin},
  R., {Dougherty}, M.~K., {Ebert}, R.~W., {Entradas}, M., {Fazakerley}, A.~N.,
  {Fortney}, J.~J., {Galand}, M., {Gustin}, J., {Hedman}, M., {Helled}, R.,
  {Henri}, P., {Hess}, S., {Holme}, R., {Karatekin}, {\"O}., {Krupp}, N.,
  {Leisner}, J., {Martin-Torres}, J., {Masters}, A., {Melin}, H., {Miller}, S.,
  {M{\"u}ller-Wodarg}, I., {Noyelles}, B., {Paranicas}, C., {de Pater}, I.,
  {P{\"a}tzold}, M., {Prang{\'e}}, R., {Qu{\'e}merais}, E., {Roussos}, E.,
  {Rymer}, A.~M., {S{\'a}nchez-Lavega}, A., {Saur}, J., {Sayanagi}, K.~M.,
  {Schenk}, P., {Schubert}, G., {Sergis}, N., {Sohl}, F., {Sittler}, E.~C.,
  {Teanby}, N.~A., {Tellmann}, S., {Turtle}, E.~P., {Vinatier}, S., {Wahlund},
  J.-E., {Zarka}, P., Sep. 2011. {Uranus Pathfinder: exploring the origins and
  evolution of Ice Giant planets}. Experimental Astronomy, 113.

\bibitem[{Bauer(1996)}]{Bauer1996}
Bauer, S.~J., 1996. Limits to the lunar ionosphere. Anzeiger Abt. II~(133),
  17--21.

\bibitem[{{Benson} et~al.(1975){Benson}, {Freeman}, {Hills}, {Ibrahim}, and
  {Schneider}}]{BFHea75}
{Benson}, J., {Freeman}, J.~W., {Hills}, H.~K., {Ibrahim}, M., {Schneider}, H.,
  Mar. 1975. {The Lunar Ionosphere}. In: Lunar and Planetary Institute
  Conference Abstracts. Vol.~6 of Lunar and Planetary Institute Conference
  Abstracts. pp. 39--+.

\bibitem[{{Bougeret} et~al.(1995){Bougeret}, {Kaiser}, {Kellogg}, {Manning},
  {Goetz}, {Monson}, {Monge}, {Friel}, {Meetre}, {Perche}, {Sitruk}, and
  {Hoang}}]{bougeret95}
{Bougeret}, J.-L., {Kaiser}, M.~L., {Kellogg}, P.~J., {Manning}, R., {Goetz},
  K., {Monson}, S.~J., {Monge}, N., {Friel}, L., {Meetre}, C.~A., {Perche}, C.,
  {Sitruk}, L., {Hoang}, S., Feb. 1995. {Waves: The Radio and Plasma Wave
  Investigation on the Wind Spacecraft}. Space Science Reviews 71, 231--263.

\bibitem[{{Bowman} and {Rogers}(2010)}]{br2010}
{Bowman}, J.~D., {Rogers}, A.~E.~E., Dec. 2010. {A lower limit of
  {$\Delta$}z$>$0.06 for the duration of the reionization epoch}. Nature 468,
  796--798.

\bibitem[{{Burns} et~al.(2012){Burns}, {Lazio}, {Bale}, {Bowman}, {Bradley},
  {Carilli}, {Furlanetto}, {Harker}, {Loeb}, and {Pritchard}}]{burns2012}
{Burns}, J.~O., {Lazio}, J., {Bale}, S., {Bowman}, J., {Bradley}, R.,
  {Carilli}, C., {Furlanetto}, S., {Harker}, G., {Loeb}, A., {Pritchard}, J.,
  Feb. 2012. {Probing the first stars and black holes in the early Universe
  with the Dark Ages Radio Explorer (DARE)}. Advances in Space Research 49,
  433--450.

\bibitem[{{Burns} et~al.(2011){Burns}, {Lazio}, {Bowman}, {Bradley}, {Carilli},
  {Furlanetto}, {Harker}, {Loeb}, and {Pritchard}}]{burns2011}
{Burns}, J.~O., {Lazio}, J., {Bowman}, J., {Bradley}, R., {Carilli}, C.,
  {Furlanetto}, S., {Harker}, G., {Loeb}, A., {Pritchard}, J., Jan. 2011. {The
  Dark Ages Radio Explorer (DARE)}. In: American Astronomical Society Meeting
  Abstracts \#217. Vol.~43 of Bulletin of the American Astronomical Society. p.
  107.09.

\bibitem[{{Burr} et~al.(2005){Burr}, {Jacobson}, and {Mielke}}]{burr05}
{Burr}, T., {Jacobson}, A., {Mielke}, A., Dec. 2005. {A dynamic global radio
  frequency noise survey as observed by the FORTE satellite at 800 km
  altitude}. Radio Science 40, 6016.

\bibitem[{{Cane} and {Erickson}(2001)}]{CaneErickson01}
{Cane}, H.~V., {Erickson}, W.~C., 2001. {A 10 MHz map of the galaxy}. Radio
  Science 36, 1765--1768.

\bibitem[{{Cane} and {Whitham}(1977)}]{CaneWhitham77}
{Cane}, H.~V., {Whitham}, P.~S., Apr. 1977. {Observations of the southern sky
  at five frequencies in the range 2-20 MHz}. Monthly Notices of the Royal
  Astronomical Society 179, 21--29.

\bibitem[{{Carilli} et~al.(2007){Carilli}, {Hewitt}, and {Loeb}}]{CHL07}
{Carilli}, C.~L., {Hewitt}, J.~N., {Loeb}, A., Feb. 2007. {Low frequency radio
  astronomy from the moon: cosmic reionization and more}. In: {Livio}, M.
  (Ed.), Proceedings of the workshop ``Astrophysics Enabled by the Return to
  the Moon''. Cambridge University Press, in press [astro-ph/0702070].

\bibitem[{{Carozzi} et~al.(2000){Carozzi}, {Karlsson}, and {Bergman}}]{CKB00}
{Carozzi}, T., {Karlsson}, R., {Bergman}, J., Feb. 2000. {Parameters
  characterizing electromagnetic wave polarization}. Physical Review E 61,
  2024--2028.

\bibitem[{{Cecconi}(2010)}]{Cecconi2010}
{Cecconi}, B., 2010. {Goniopolarimetric techniques for low-frequency radio
  astronomy in space}. ISSI Scientific Reports Series 9, 263--277.

\bibitem[{{Chapman} et~al.(2012){Chapman}, {Abdalla}, {Harker}, {Jeli{\'c}},
  {Labropoulos}, {Zaroubi}, {Brentjens}, {de Bruyn}, and {Koopmans}}]{chap2012}
{Chapman}, E., {Abdalla}, F.~B., {Harker}, G., {Jeli{\'c}}, V., {Labropoulos},
  P., {Zaroubi}, S., {Brentjens}, M.~A., {de Bruyn}, A.~G., {Koopmans},
  L.~V.~E., Jul. 2012. {Foreground removal using FASTICA: a showcase of
  LOFAR-EoR}. Monthly Notices of the Royal Astronomical Society 423,
  2518--2532.

\bibitem[{{Christophe} et~al.(2011){Christophe}, {Spilker}, {Anderson},
  {Andr{\'e}}, {Asmar}, {Aurnou}, {Banfield}, {Barucci}, {Bertolami},
  {Bingham}, {Brown}, {Cecconi}, {Courty}, {Dittus}, {Fletcher}, {Foulon},
  {Francisco}, {Gil}, {Glassmeier}, {Grundy}, {Hansen}, {Helbert}, {Helled},
  {Hussmann}, {Lamine}, {L{\"a}mmerzahl}, {Lamy}, {Lenoir}, {Levy}, {Orton},
  {P{\'a}ramos}, {Poncy}, {Postberg}, {Progrebenko}, {Reh}, {Reynaud},
  {Robert}, {Samain}, {Saur}, {Sayanagi}, {Schmitz}, {Selig}, {Sohl},
  {Spilker}, {Srama}, {Stephan}, {Touboul}, and {Wolf}}]{Chris2011}
{Christophe}, B., {Spilker}, L.~J., {Anderson}, J.~D., {Andr{\'e}}, N.,
  {Asmar}, S.~W., {Aurnou}, J., {Banfield}, D., {Barucci}, A., {Bertolami}, O.,
  {Bingham}, R., {Brown}, P., {Cecconi}, B., {Courty}, J.-M., {Dittus}, H.,
  {Fletcher}, L.~N., {Foulon}, B., {Francisco}, F., {Gil}, P.~J.~S.,
  {Glassmeier}, K.-H., {Grundy}, W., {Hansen}, C., {Helbert}, J., {Helled}, R.,
  {Hussmann}, H., {Lamine}, B., {L{\"a}mmerzahl}, C., {Lamy}, L., {Lenoir}, B.,
  {Levy}, A., {Orton}, G., {P{\'a}ramos}, J., {Poncy}, J., {Postberg}, F.,
  {Progrebenko}, S.~V., {Reh}, K.~R., {Reynaud}, S., {Robert}, C., {Samain},
  E., {Saur}, J., {Sayanagi}, K.~M., {Schmitz}, N., {Selig}, H., {Sohl}, F.,
  {Spilker}, T.~R., {Srama}, R., {Stephan}, K., {Touboul}, P., {Wolf}, P., Jun.
  2011. {OSS (Outer Solar System): A fundamental and planetary physics mission
  to Neptune, Triton and the Kuiper Belt}. ArXiv e-prints.

\bibitem[{{Ciardi} and {Ferrara}(2005)}]{CiardiFerrara05}
{Ciardi}, B., {Ferrara}, A., Feb. 2005. {The First Cosmic Structures and Their
  Effects}. Space Science Reviews 116, 625--705, as updated at
  arXiv:astro-ph/0409018.

\bibitem[{{Ciardi} and {Salvaterra}(2007)}]{CS07}
{Ciardi}, B., {Salvaterra}, R., Nov. 2007. {Ly{$\alpha$} heating and its impact
  on early structure formation}. Monthly Notices of the Royal Astronomical
  Society 381, 1137--1142.

\bibitem[{{Close} et~al.(2010){Close}, {Colestock}, {Cox}, {Kelley}, and
  {Lee}}]{close2010}
{Close}, S., {Colestock}, P., {Cox}, L., {Kelley}, M., {Lee}, N., Dec. 2010.
  {Electromagnetic pulses generated by meteoroid impacts on spacecraft}.
  Journal of Geophysical Research (Space Physics) 115~(A14), 12328.

\bibitem[{{Ellis} and {Mendillo}(1987)}]{EllisMendillo87}
{Ellis}, G.~R.~A., {Mendillo}, M., 1987. {A 1.6 MHz survey of the galactic
  background radio emission}. Australian Journal of Physics 40, 705--708.

\bibitem[{{Fridman} and {Baan}(2001)}]{fridman01}
{Fridman}, P.~A., {Baan}, W.~A., Oct. 2001. {RFI mitigation methods in radio
  astronomy}. Astronomy and Astrophysics 378, 327--344.

\bibitem[{{Furlanetto} et~al.(2009){Furlanetto}, {Lidz}, {Loeb}, {McQuinn},
  {Pritchard}, {Shapiro}, {Alvarez}, {Backer}, {Bowman}, {Burns}, {Carilli},
  {Cen}, {Cooray}, {Gnedin}, {Greenhill}, {Haiman}, {Hewitt}, {Hirata},
  {Lazio}, {Mesinger}, {Madau}, {Morales}, {Oh}, {Peterson}, {Pihlstr{\"o}m},
  {Tegmark}, {Trac}, {Zahn}, and {Zaldarriaga}}]{fur2009}
{Furlanetto}, S.~R., {Lidz}, A., {Loeb}, A., {McQuinn}, M., {Pritchard}, J.~R.,
  {Shapiro}, P.~R., {Alvarez}, M.~A., {Backer}, D.~C., {Bowman}, J.~D.,
  {Burns}, J.~O., {Carilli}, C.~L., {Cen}, R., {Cooray}, A., {Gnedin}, N.,
  {Greenhill}, L.~J., {Haiman}, Z., {Hewitt}, J.~N., {Hirata}, C.~M., {Lazio},
  J., {Mesinger}, A., {Madau}, P., {Morales}, M.~F., {Oh}, S.~P., {Peterson},
  J.~B., {Pihlstr{\"o}m}, Y.~M., {Tegmark}, M., {Trac}, H., {Zahn}, O.,
  {Zaldarriaga}, M., 2009. {Cosmology from the Highly-Redshifted 21 cm Line}.
  In: astro2010: The Astronomy and Astrophysics Decadal Survey. Vol. 2010 of
  ArXiv Astrophysics e-prints. p.~82.

\bibitem[{{Furlanetto} et~al.(2006){Furlanetto}, {Oh}, and {Briggs}}]{FOB06}
{Furlanetto}, S.~R., {Oh}, S.~P., {Briggs}, F.~H., Oct. 2006. {Cosmology at low
  frequencies: The 21 cm transition and the high-redshift Universe}. Physics
  Reports 433, 181--301.

\bibitem[{{Gardini}(2011)}]{gardini2011}
{Gardini}, B., 2011. {ESA strategy for human exploration and the Lunar Lander
  Mission}. Memorie della Societa Astronomica Italiana 82, 422.

\bibitem[{{Genova} et~al.(1989){Genova}, {Zarka}, and {Lecacheux}}]{GZL1989}
{Genova}, F., {Zarka}, P., {Lecacheux}, A., 1989. {Jupiter Decametric
  Radiation}. NASA Special Publication 494, 156--174.

\bibitem[{{Goto} et~al.(2011){Goto}, {Fujimoto}, {Kasahara}, {Kumamoto}, and
  {Ono}}]{goto11}
{Goto}, Y., {Fujimoto}, T., {Kasahara}, Y., {Kumamoto}, A., {Ono}, T., Jan.
  2011. {Lunar ionosphere exploration method using auroral kilometric
  radiation}. Earth, Planets, and Space 63, 47--56.

\bibitem[{{Grie{\ss}meier} et~al.(2011){Grie{\ss}meier}, {Zarka}, and
  {Girard}}]{GZG2011}
{Grie{\ss}meier}, P., {Zarka}, P., {Girard}, J.~N., 2011. {Observation of
  planetary radio emissions using large arrays}. RADIO SCIENCE 46, 4752.

\bibitem[{{Harker} et~al.(2009){Harker}, {Zaroubi}, {Bernardi}, {Brentjens},
  {de Bruyn}, {Ciardi}, {Jeli{\'c}}, {Koopmans}, {Labropoulos}, {Mellema},
  {Offringa}, {Pandey}, {Schaye}, {Thomas}, and {Yatawatta}}]{harker2009}
{Harker}, G., {Zaroubi}, S., {Bernardi}, G., {Brentjens}, M.~A., {de Bruyn},
  A.~G., {Ciardi}, B., {Jeli{\'c}}, V., {Koopmans}, L.~V.~E., {Labropoulos},
  P., {Mellema}, G., {Offringa}, A., {Pandey}, V.~N., {Schaye}, J., {Thomas},
  R.~M., {Yatawatta}, S., Aug. 2009. {Non-parametric foreground subtraction for
  21-cm epoch of reionization experiments}. Monthly Notices of the Royal
  Astronomical Society 397, 1138--1152.

\bibitem[{{Harker} et~al.(2012){Harker}, {Pritchard}, {Burns}, and
  {Bowman}}]{harker2012}
{Harker}, G.~J.~A., {Pritchard}, J.~R., {Burns}, J.~O., {Bowman}, J.~D., Jan.
  2012. {An MCMC approach to extracting the global 21-cm signal during the
  cosmic dawn from sky-averaged radio observations}. Monthly Notices of the
  Royal Astronomical Society 419, 1070--1084.

\bibitem[{{Imamura} et~al.(2008){Imamura}, {Iwata}, {Yamamoto}, {Oyama},
  {Nabatov}, {Kono}, {Matsumoto}, {Liu}, {Noda}, {Hanada}, {Futaana}, and
  {Saito}}]{ima2008}
{Imamura}, T., {Iwata}, T., {Yamamoto}, Z., {Oyama}, K.-I., {Nabatov}, A.,
  {Kono}, Y., {Matsumoto}, M., {Liu}, Q., {Noda}, H., {Hanada}, H., {Futaana},
  Y., {Saito}, A., Mar. 2008. {Initial Results of the Lunar Ionosphere
  Observation with SELENE Radio Science}. In: Lunar and Planetary Institute
  Science Conference Abstracts. Vol.~39 of Lunar and Planetary Institute
  Science Conference Abstracts. p. 1659.

\bibitem[{{Jester} and {Falcke}(2009)}]{JF2009}
{Jester}, S., {Falcke}, H., May 2009. {Science with a lunar low-frequency
  array: From the dark ages of the Universe to nearby exoplanets}. New
  Astronomy Review 53, 1--26.

\bibitem[{{Kaiser} et~al.(1996){Kaiser}, {Desch}, {Bougeret}, {Manning}, and
  {Meetre}}]{kaiser96}
{Kaiser}, M.~L., {Desch}, M.~D., {Bougeret}, J.-L., {Manning}, R., {Meetre},
  C.~A., 1996. {Wind/WAVES observations of man-made radio transmissions}.
  Geophysical Research Letters 23, 1287--1290.

\bibitem[{{Kassim} and {Weiler}(1990)}]{KassimWeiler90}
{Kassim}, N.~E., {Weiler}, K.~W. (Eds.), 1990. {Low frequency astrophysics from
  space; Proceedings of an International Workshop, Crystal City, VA, Jan. 8, 9,
  1990}.

\bibitem[{{Lamy}(2011)}]{lamy2011}
{Lamy}, L., 2011. {Variability of southern and northern periodicities of Saturn
  Kilometric Radiation}. Planetary, Solar and Heliospheric Radio Emissions (PRE
  VII), 38--50.

\bibitem[{{Lamy} et~al.(2012){Lamy}, {Prangé}, {Hansen}, {Clarke}, {Zarka},
  and {Cecconi}}]{Lamy2012}
{Lamy}, L., {Prangé}, R., {Hansen}, K.~C., {Clarke}, J.~T., {Zarka}, P.,
  {Cecconi}, B. e.~a., 2012. {Earth-based detection of Uranus' aurorae}.
  Geophysics Research Letters, in press.

\bibitem[{{Lazio} et~al.(2011){Lazio}, {MacDowall}, {Burns}, {Jones}, {Weiler},
  {Demaio}, {Cohen}, {Paravastu Dalal}, {Polisensky}, {Stewart}, {Bale},
  {Gopalswamy}, {Kaiser}, and {Kasper}}]{lazio2011}
{Lazio}, T.~J.~W., {MacDowall}, R.~J., {Burns}, J.~O., {Jones}, D.~L.,
  {Weiler}, K.~W., {Demaio}, L., {Cohen}, A., {Paravastu Dalal}, N.,
  {Polisensky}, E., {Stewart}, K., {Bale}, S., {Gopalswamy}, N., {Kaiser}, M.,
  {Kasper}, J., Dec. 2011. {The Radio Observatory on the Lunar Surface for
  Solar studies}. Advances in Space Research 48, 1942--1957.

\bibitem[{{Mann} et~al.(1999){Mann}, {Jansen}, {MacDowall}, {Kaiser}, and
  {Stone}}]{Mann99}
{Mann}, G., {Jansen}, F., {MacDowall}, R.~J., {Kaiser}, M.~L., {Stone}, R.~G.,
  Aug. 1999. {A heliospheric density model and type III radio bursts}.
  Astronomy and Astrophysics 348, 614--620.

\bibitem[{{Novaco} and {Brown}(1978)}]{NB78}
{Novaco}, J.~C., {Brown}, L.~W., Apr. 1978. {Nonthermal galactic emission below
  10 megahertz}. Astrophysical Journal 221, 114--123.

\bibitem[{{Ogilvie} and {Parks}(1996)}]{op96}
{Ogilvie}, K.~W., {Parks}, G.~K., 1996. {First results from WIND spacecraft: An
  introduction}. Geophysics Research Letters 23, 1179--1182.

\bibitem[{{Pluchino} et~al.(2008){Pluchino}, {Schillir{\`o}}, {Salerno},
  {Pupillo}, {Maccaferri}, and {Cassaro}}]{pluch2008}
{Pluchino}, S., {Schillir{\`o}}, F., {Salerno}, E., {Pupillo}, G.,
  {Maccaferri}, G., {Cassaro}, P., 2008. {Radio occultation measurements of the
  lunar ionosphere.} Memorie della Societa Astronomica Italiana Supplementi 12,
  53.

\bibitem[{{Pritchard} and {Loeb}(2010)}]{prit2010}
{Pritchard}, J., {Loeb}, A., Dec. 2010. {Cosmology: Hydrogen was not ionized
  abruptly}. Nature 468, 772--773.

\bibitem[{{Reasoner} and {Burke}(1972)}]{RB72}
{Reasoner}, D.~L., {Burke}, W.~J., 1972. {Direct observation of the lunar
  photoelectron layer}. In: {Metzger}, A.~E., {Trombka}, J.~I., {Peterson},
  L.~E., {Reedy}, R.~C., {Arnold}, J.~R. (Eds.), Lunar and Planetary Science
  Conference. Vol.~3 of Lunar and Planetary Science Conference. pp. 2639--+.

\bibitem[{{Reasoner} and {O'Brien}(1972)}]{RO72}
{Reasoner}, D.~L., {O'Brien}, B.~J., 1972. {Measurement on the lunar surface of
  impact-produced plasma clouds.} Journal of Geophysical Research 77,
  1292--1299.

\bibitem[{{Rees}(1999)}]{Rees99}
{Rees}, M.~J., May 1999. {The End of the `Dark Age'}. In: {Holt}, S., {Smith},
  E. (Eds.), After the Dark Ages: When Galaxies were Young (the Universe at $2
  < z < 5$). Vol. 470 of American Institute of Physics Conference Series. pp.
  13--+.

\bibitem[{{Scholten} et~al.(2006){Scholten}, {Bacelar}, {Braun}, {de Bruyn},
  {Falcke}, {Stappers}, and {Strom}}]{scholten06}
{Scholten}, O., {Bacelar}, J., {Braun}, R., {de Bruyn}, A.~G., {Falcke}, H.,
  {Stappers}, B., {Strom}, R.~G., Oct. 2006. {Optimal radio window for the
  detection of Ultra-High Energy cosmic rays and neutrinos off the moon}.
  Astroparticle Physics 26, 219--229.

\bibitem[{{St{\aa}l} et~al.(2007){St{\aa}l}, {Bergman}, {Thid{\'e}},
  {Daldorff}, and {Ingelman}}]{staal2007}
{St{\aa}l}, O., {Bergman}, J.~E.~S., {Thid{\'e}}, B., {Daldorff}, L.~K.~S.,
  {Ingelman}, G., Feb. 2007. {Prospects for Lunar Satellite Detection of Radio
  Pulses from Ultrahigh Energy Neutrinos Interacting with the Moon}. Physical
  Review Letters 98~(7), 071103.

\bibitem[{{Stubbs} et~al.(2007){Stubbs}, {Vondrak}, and {Farrell}}]{stubbs2007}
{Stubbs}, T.~J., {Vondrak}, R.~R., {Farrell}, W.~M., Jan. 2007. {Impact of Dust
  on Lunar Exploration}. Dust in Planetary Systems 643, 239--243.

\bibitem[{{Takahashi}(2003)}]{taka03}
{Takahashi}, Y.~D., Jun. 2003. {A concept for a simple radio observatory at the
  lunar south pole}. Advances in Space Research 31, 2473--2478.

\bibitem[{{Vyshlov}(1976)}]{Vyshlov1976}
{Vyshlov}, A.~S., 1976. {Preliminary results of circumlunar plasma research by
  the Luna 22 spacecraft}, 945--949.

\bibitem[{{Weiler}(1987)}]{Weiler87}
{Weiler}, K.~W. (Ed.), 1987. {Radio astronomy from space; Proceedings of the
  Workshop, Green Bank, WV, Sept. 30-Oct. 2, 1986}.

\bibitem[{{Weiler} et~al.(1988){Weiler}, {Johnston}, {Simon}, {Dennison},
  {Erickson}, {Kaiser}, {Cane}, and {Desch}}]{WJSea88}
{Weiler}, K.~W., {Johnston}, K.~J., {Simon}, R.~S., {Dennison}, B.~K.,
  {Erickson}, W.~C., {Kaiser}, M.~L., {Cane}, H.~V., {Desch}, M.~D., Apr. 1988.
  {A low frequency radio array for space}. Astronomy and Astrophysics 195,
  372--379.

\bibitem[{{Well} and {Barasch}(1962)}]{well1962}
{Well}, H., {Barasch}, M.~L., Jun. 1962. {Model of the Lunar Ionosphere.}
  Astronomical Journal 67, 589.

\bibitem[{{Zarka}(1998)}]{Zarka1998}
{Zarka}, P., Sep. 1998. {Auroral radio emissions at the outer planets:
  Observations and theories}. Journal of Geophysical Research 103,
  20159--20194.

\bibitem[{{Zarka}(2007)}]{Zarka2007}
{Zarka}, P., Apr. 2007. {Plasma interactions of exoplanets with their parent
  star and associated radio emissions}. Planetary Space Science 55, 598--617.

\bibitem[{{Zarka} et~al.(2004){Zarka}, {Cecconi}, and {Kurth}}]{Zarka2004}
{Zarka}, P., {Cecconi}, B., {Kurth}, W.~S., Aug. 2004. {Jupiter's low-frequency
  radio spectrum from Cassini/Radio and Plasma Wave Science (RPWS) absolute
  flux density measurements}. Journal of Geophysical Research (Space Physics)
  109~(A18), 9.

\bibitem[{{Zarka} et~al.(2008){Zarka}, {Farrell}, {Fischer}, and
  {Konovalenko}}]{Zarka2008}
{Zarka}, P., {Farrell}, W., {Fischer}, G., {Konovalenko}, A., Jun. 2008.
  {Ground-Based and Space-Based Radio Observations of Planetary Lightning}.
  Space Science Reviews 137, 257--269.

\end{thebibliography}
\end{document}